\documentclass[a4paper,12pt]{article}
\pdfoutput=1
\usepackage[a4paper,text={16.8cm,22.4cm}]{geometry}
\usepackage{amsmath,amssymb,bm,braket,psfrag,colortbl,booktabs,latexsym}
\usepackage{graphicx}
\usepackage{color}
\usepackage{cite}

\allowdisplaybreaks 
\newcommand{\Tr}{\mathrm{Tr}}
\newcommand{\Li}{\mathrm{Li}}
\newcommand{\ff}{f\hspace{-0.4em}f}

\newcommand{\IPIs}{\mbox{1PI}_{\mbox{\tiny{SCET}}}}
\newcommand{\PIMs}{\mbox{PIM}_{\mbox{\tiny{SCET}}}}

\newcommand{\be}{\begin{equation}}
\newcommand{\ee}{\end{equation}}
\newcommand{\st}{\tilde{t}_1}

\begin{document}

\begin{titlepage}

  \begin{flushright}
    {MITP/13-10}\\
    {PSI-PR-13-18}\\
    April 8, 2013
  \end{flushright}
  
  \vspace{5ex}
  
  \begin{center}
    \textbf{\Large Approximate NNLO Predictions for the Stop-Pair Production Cross Section at the LHC} \vspace{7ex}
    
    \textsc{Alessandro Broggio$^{a,b}$, Andrea Ferroglia$^c$, Matthias Neubert$^{a}$,\\
     Leonardo Vernazza$^{a,d}$ and Li Lin Yang$^{e,f,g}$} \vspace{2ex}
  
    \textsl{${}^a$PRISMA Cluster of Excellence \& Mainz Institut for Theoretical Physics\\
      Johannes Gutenberg University, D-55099 Mainz, Germany\\[0.3cm]
      ${}^b$Paul Scherrer Institute\\ CH-5232 Villigen, Switzerland\\[0.3cm]
      ${}^c$New York City College of Technology\\ 300 Jay Street,
            Brooklyn, NY 11201, USA\\[0.3cm]
      ${}^d$Dipartimento di Fisica, Universit\`a di Torino \& INFN, Sezione di Torino\\
      Via P. Giuria 1, I-10125 Torino, Italy\\[0.3cm]
      ${}^e$Institute for Theoretical Physics\\ University of Zurich, 
      CH-8057 Zurich, Switzerland\\[0.3cm]
      ${}^f$School of Physics and State Key Laboratory of Nuclear Physics and Technology\\
      Peking University, 100871 Beijing, China\\[0.3cm]
      ${}^g$Center for High Energy Physics, Peking University, Beijing 100871, China}
  \end{center}

\vspace{4ex}

\begin{abstract}
If the minimal supersymmetric standard model at scales of around $1~$TeV is realized in nature, 
the total top-squark pair production cross section should be measurable at the CERN Large Hadron Collider. In this work we present precise predictions for this observable, which are based upon approximate NNLO formulas obtained using soft-collinear effective theory methods.
\end{abstract}

\end{titlepage}

\section{Introduction}

One of the main goals of the physics program at the Large Hadron Collider (LHC) is to investigate the existence of supersymmetric partners of the fundamental particles with masses in the TeV range. Within the context of the minimal supersymmetric standard model (MSSM) with R-parity conservation, supersymmetric particles are produced in pairs. At hadron colliders, the supersymmetric particles which are expected to be most abundantly produced are the ones which carry color charge: squarks and gluinos. Since supersymmetry is broken, the mass spectrum of squarks and gluinos plays a crucial part in determining which among the colored supersymmetric partners is  experimentally accessible.
Within the context of unified supersymmetric theories, the scalar and gaugino masses are evolved from a common high scale down to the energy scale of electroweak symmetry breaking. As a consequence of large Yukawa and soft couplings entering the evolution of the mass parameters, the third generation of squarks can have very different masses compared to the first two generations of squarks.
In particular, under the assumption of ``natural supersymmetry'' \cite{Papucci:2011wy}, top-squarks are expected to be relatively light in order to reproduce the correct energy scale of electroweak symmetry breaking.
Therefore, the lightest of the two supersymmetric partners of the top quark is usually expected to be the lightest squark in the  mass spectrum.
Precise theoretical predictions of the stop pair production cross section are instrumental in setting a lower bound on the lightest stop mass. Moreover, if top squarks will be discovered, accurate predictions of the cross section for stop pair production can be employed to determine the masses and other properties of these particles.
For these reasons, the study of the radiative corrections to the production of stop pairs has already a quite long history.
The calculation of the next-to-leading order (NLO) corrections to the cross section for stop pair production within the context of supersymmetric quantum chromodynamics (SUSY-QCD) was completed 15 years ago \cite{Beenakker:1997ut}. As expected, it was found that the NLO corrections significantly decrease the renormalization- and factorization-scale dependences of the prediction when compared to the leading order (LO) calculation.  Furthermore, NLO SUSY-QCD corrections increase the value of the cross section if the renormalization and factorization scales are chosen close to the value of the stop mass.
The NLO SUSY-QCD corrections are implemented in the computer programs {\tt Prospino} and {\tt Prospino2} \cite{Beenakker:1996ed}.
The electroweak corrections to stop pair production were studied in \cite{Hollik:2007wf, Beccaria:2008mi}. While these corrections have a quite sizable effect on the 
tails of the invariant-mass and transverse-momentum distributions, they only have a moderate impact on the total cross section. 
The emission of  soft gluons accounts for a significant portion of the NLO SUSY-QCD corrections \cite{Beenakker:1996ch}, which are large. For this reason, the resummation of the next-to-leading logarithmic (NLL) corrections was carried out in \cite{Beenakker:2010nq}. It was found that these  corrections increase the cross section at the LHC by up to $10 \%$ of its NLO value, while they further decrease the scale dependence of the  prediction.  In \cite{Langenfeld:2009eg,Langenfeld:2010vu,Langenfeld:2012ti} 
next-to-next-to-leading order (NNLO) threshold corrections and Coulomb corrections were derived by means of resummation techniques.  In these studies,  the resummation is carried out in Mellin moment space.

In the last few years, a formalism based on soft-collinear effective theory (SCET), which allows one to resum soft-gluon corrections directly in momentum space, was developed \cite{Becher:2006mr,Becher:2006nr} and applied to QCD corrections for several processes of interest in collider phenomenology, such as Drell-Yan scattering \cite{Becher:2007ty}, Higgs production \cite{Ahrens:2008nc, Ahrens:2008qu}, direct photon production \cite{Becher:2009th}, and recently slepton pair production \cite{Broggio:2011bd}. A similar approach was developed independently in \cite{Beneke:2010da}, where methods of SCET and non-relativistic QCD were used to resum simultaneously the effects of soft and Coulomb gluons. This method was applied to squark and gluino pair production \cite{Beneke:2010da,Falgari:2012hx}, where soft and Coulomb-gluon contributions were resummed at NLL order, and to top-quark pair production \cite{Beneke:2011mq}, where resummation up to next-to-next-to-leading logarithmic (NNLL) order was implemented. 
The soft-gluon corrections to the top-quark pair production cross section at the Tevatron and the LHC were also studied in \cite{Ahrens:2010zv, Ahrens:2011mw}, within the framework developed in \cite{Becher:2006mr,Becher:2006nr}. By employing this formalism, it was possible to resum soft-gluon emission corrections up to NNLL accuracy, as well as to derive approximate NNLO formulas for the total production cross section, the pair invariant-mass distribution, the top-quark transverse-momentum spectrum, and the top-quark rapidity distribution.

In fixed-order perturbation theory, differential partonic cross sections involve both singular distributions and regular terms.
The singular distributions are functions of a ``soft parameter'', whose precise definition depends on the kinematics.
In \cite{Ahrens:2010zv,Becher:2007ty}, where pair-invariant mass (PIM) kinematics was employed, the relevant soft parameter is $\sqrt{s}(1-z)$, where $\sqrt{s}$ is the partonic center of mass energy, $z = M^2/ s$, and $M$ is the pair invariant mass of the energetic particles produced.
In this case, the soft (or partonic-threshold) region corresponds to the limit $z \to 1$.
In \cite{Ahrens:2011mw}, where one-particle inclusive (1PI) kinematics was employed, the relevant soft parameter is $s_4$, which is obtained by subtracting the heavy-particle mass squared from the invariant mass of the objects recoiling against the observed heavy particle. In this case, the soft gluon emission region is identified by taking the limit $s_4 \to 0$.
Detailed studies of Drell-Yan and top pair production processes showed that, after the convolution of the hard-scattering kernels
with the parton luminosities, the terms which are singular  in the partonic-threshold limit provide the numerically  dominant contributions to the hadronic cross sections. In these processes, the singular terms  typically account for more than 90\% of the total NLO corrections.
Furthermore, the dominance of the partonic-threshold regions in the
calculation of the cross sections arises dynamically, since it is due to the strong fall-off of the parton
luminosities outside the threshold region.
This phenomenon is called ``dynamical threshold enhancement'' \cite{Becher:2007ty}. 

In the partonic-threshold region, the partonic cross sections factorize into the product of hard functions, which contain the virtual corrections to the cross section, and soft functions, which account for the effects of soft gluon emissions.
The singular (plus) distributions in the partonic cross section can be obtained by solving the renormalization group equations (RGEs) satisfied by the hard and soft functions. The corresponding anomalous dimensions are known up to NNLO \cite{Ferroglia:2009ep,Ferroglia:2009ii}. The information extracted from the RGE and the knowledge of the hard and soft functions at NLO allows us to obtain approximate formulas including all of the singular terms up to NNLO. These approximate NNLO formulas can then be matched with exact NLO calculations in order to obtain precise theoretical predictions for the observables of interest.

Within this context, the production of top-squark pairs in the soft-gluon emission limit can be studied in analogy to the production of top-quark pairs.
The hard and soft functions are matrices in color space. The soft functions in the stop-quark and top-quark production processes are identical; they were evaluated in \cite{Ahrens:2010zv, Ahrens:2011mw} up to NLO. In contrast,  the hard functions for the stop-pair production process are so far unknown and must be computed with NLO accuracy. It is important to observe that the only SUSY parameter which appears in the soft functions is the mass of the produced stop quark, while all of the other SUSY parameters appear exclusively in the hard functions.
In this work, we carry out the calculation of the hard functions for top-squark pair production up to NLO. By combining the NLO hard functions with the 
NLO soft functions and with the anomalous dimensions entering the 
RGEs satisfied by the various terms in the factorized cross section, it is possible to resum soft-gluon emission corrections up to NNLL order. Here we limit ourselves to re-expand the resummed formulas in order to obtain approximate NNLO formulas
for the pair invariant-mass spectrum and the stop transverse-momentum and rapidity distributions. 
Although our formulas enable us to obtain predictions for these differential distributions, we focus our attention on the total top-squark production cross section, which is the observable of most immediate interest in the top squark searches.
In fact, by integrating the approximate NNLO formulas for the differential distributions over the complete phase-space, we obtain predictions for the total top-squark pair production cross section at the LHC, and we comment on the phenomenological impact of the NNLO corrections arising from soft emissions.
In a future work, we will evaluate the resummed stop pair production cross section and discuss the phenomenological impact of soft gluon resummation at NNLL accuracy. 
It must be remarked that the resummation  at NNLL  accuracy, in Mellin space, is already available for the production of other SUSY particles; the production of squark pairs was carried out  in \cite{Beenakker:2011sf}, while the production of gluino pairs was studied at this level of accuracy in \cite{Pfoh:2013iia}. Furthermore the NLO hard matching
coefficients for squark and gluino hadroproduction, an important ingredient for NNLL studies,  were evaluated very recently \cite{Beenakker:2013mva}.

The paper is organized as follows: In Section~\ref{sec:Notation} we introduce our notation and conventions, which are very similar to the ones employed in 
\cite{Ahrens:2010zv, Ahrens:2011mw}. Furthermore, we describe the factorization of the stop pair production cross section in the soft limit, 
both in PIM kinematics and in 1PI kinematics. In Section~\ref{sec:Matching}, we discuss the calculation of the soft and hard functions up to NLO. 
In Section~\ref{sec:RGEeq}, we discuss the structure of  the hard scattering kernels  and we present approximate NNLO formulas for the stop pair production process. Predictions for the total top-squark pair production cross section at the LHC can be found in Section~\ref{sec:CS}, together with an analysis of the phenomenological impact of the approximate NNLO corrections on this observable. 
 We conclude Section~\ref{sec:CS} by comparing our results with the NLO+NLL results of and 
\cite{Beenakker:2010nq} and
\cite{Falgari:2012hx}, which are obtained by means of techniques and kinematic schemes different from the ones that we employ in this work.
Finally, we collect our conclusions in Section~\ref{sec:Conc}.

\section{Notation}
\label{sec:Notation}

\begin{figure}[t]
\begin{center}
\begin{tabular}{c} \includegraphics[width=0.25\textwidth]{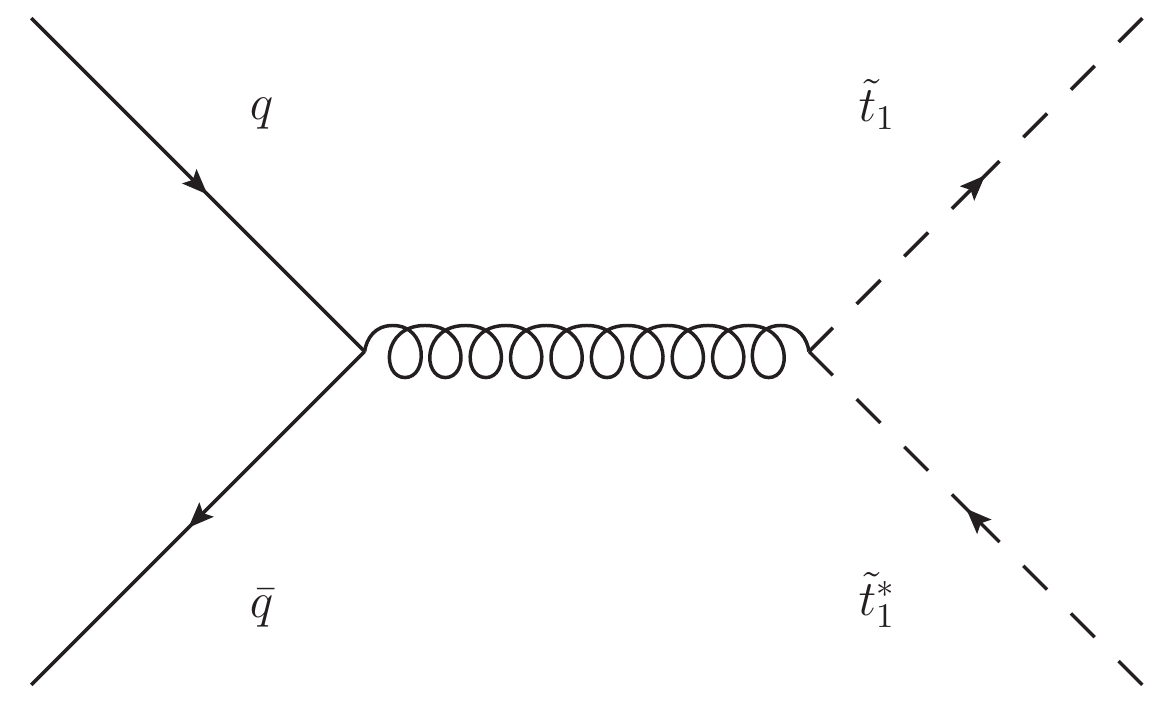}\\ 
 \\
 \includegraphics[width=0.98\textwidth]{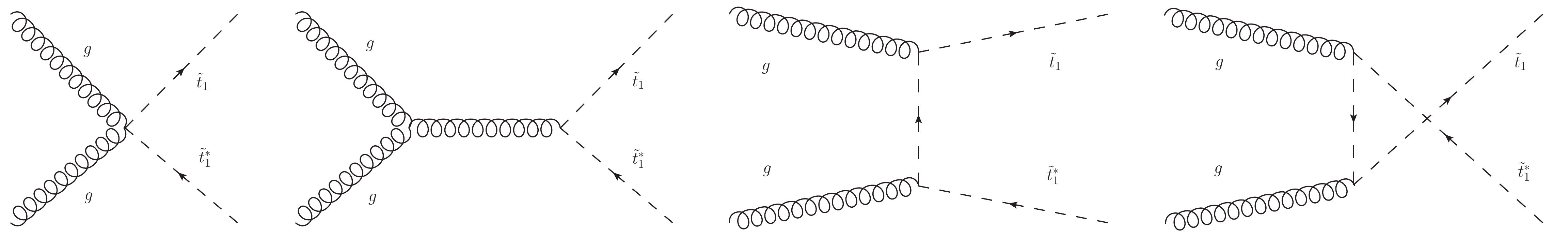}\\ 
\end{tabular} 
\end{center}
\caption{Tree level diagrams contributing to the quark annihilation production channel (first row) and gluon fusion channel (second row).
\label{TLfig}}
\end{figure}

In this paper we study the process
\begin{align}
N_1(P_1) + N_2(P_2) &\to \st(p_3) + \st^*(p_4) + X(k) \, , \label{eq:hadproc}
\end{align}
where $N_1$ and $N_2$ indicate the incoming protons in the case of a proton-proton collider as the LHC, while $X$ is an inclusive hadronic final state. In the rest of the paper, we treat the top squarks as on-shell particles and neglect their decay. The terms neglected in this approximation are of order $\Gamma_{\st}/m_{\st}$.  At the lowest order in perturbation theory, two partonic channels contribute to the process in Eq.~(\ref{eq:hadproc}):
\begin{align}
q(p_1) + \bar{q}(p_2) &\to \st(p_3) + \st^*(p_4) \, , \nonumber \\
g(p_1) + g(p_2) &\to  \st(p_3) + \st^*(p_4) \, ,
\end{align}
where the momenta of the incoming partons are related to the momenta of the incoming hadrons by $p_i = x_i P_i$ ($i=1,2$).
The diagrams contributing to the two production channels at lowest order in QCD are shown in Figure~\ref{TLfig}.
The relevant invariants for the hadronic scattering are 
\begin{align}
S = (P_1+P_2)^2 \,  ,  \quad T_1 = (P_1 -p_3)^2 - m_{\st}^2 \, , \quad U_1 = (P_1 - p_4)^2 - m_{\st}^2 \, .
\end{align}
In order to describe the partonic scattering, we employ the invariants
\begin{align}
s = x_1 x_2 S = (p_1 +p_2)^2 \, , \quad t_1 = x_1 T_1 \, , \quad u_1 = x_2 U_1 \, , \nonumber \\
M^2 = (p_3 + p_4)^2 \, , \quad  s_4 = s+t_1 +u_1 = (p_4+k)^2 - m_{\st}^2 \, . \label{eq:ManPart}
\end{align}
In Born approximation $s+t_1+u_1=0$, and consequently $M^2 = s$ and $s_4 = 0$.

It is well known that the kinematics of the process allows one to define different threshold regions. Here we consider two different cases: PIM
kinematics, in which the threshold region is defined by the limit $s \to M^2$, and 1PI kinematics, in which the threshold region is 
approached by taking the limit $s_4 \to 0$. Both regions were employed in the study of the top-quark pair production cross section and differential distributions \cite{Kidonakis:2001nj,Ahrens:2009uz,Ahrens:2010zv,Ahrens:2011mw}.
In particular, by working in PIM kinematics one can calculate the pair invariant-mass distribution, while by working in 1PI kinematics
one can evaluate the stop transverse-momentum and rapidity distributions.
It should be emphasized that in the PIM and 1PI threshold regions the top squarks are not forced to be nearly at rest, as in case of the threshold region defined by the limit $\beta = \sqrt{1- 4 m_{\st}/s } \to 0$, which is often employed in the calculation of soft-gluon corrections to the total cross section \cite{Beenakker:2010nq,Beneke:2010da,Falgari:2012hx}. In the rest of the paper, we refer to the $\beta \to 0$ limit as the production threshold region.

Our goal is to employ both  PIM and 1PI kinematics to obtain approximate NNLO formulas for the total top-squark pair production cross section.
Both approaches include the numerically large contributions arising from the emission of soft gluons. Differences between them, and with the production threshold calculations, arise from the way in which different sets of power-suppressed corrections are treated. 

\subsection{PIM kinematics}

We focus first on the case of PIM kinematics. It is convenient to introduce the quantities
\be
z = \frac{M^2}{s} \, , \qquad \tau = \frac{M^2}{S} \, , \qquad \beta_{\st} = \sqrt{1- \frac{4 m_{\st}^2}{M^2}} \, .
\ee
Consequently, the PIM threshold limit $s \to M^2$ corresponds to the limit $z \to 1$.
According to the QCD factorization theorem \cite{Collins:1989gx}, the differential cross section  in $M$  and $\theta$ (the scattering angle of the top squark with respect to the beam axis in the partonic rest frame) is given by
\be
\frac{d^2 \sigma}{d M d \cos{\theta}} = \frac{\pi \beta_{\st}}{S M} \sum_{i,j} \int_\tau^1 \frac{dz}{z} \ff_{ij}\left(\frac{\tau}{z}, \mu_f\right) C_{{\rm PIM}, ij}\left(z,M, \cos{\theta}, \mu_f\right) ,
\label{eq:dMdcos}
\ee
where $\mu_f$ is the factorization scale, and the sum runs over the incoming partons. 
In the following we drop the subscript PIM (and the corresponding subscript 1PI) whenever there is no ambiguity about the kinematic scheme employed, or when a formula applies to both schemes.
The parton luminosities $\ff_{ij}$ are defined as the convolutions of the 
non-perturbative parton distribution functions (PDF) for the incoming partons $i$ and $j$:
\be
\ff_{ij}(y,\mu_f) = \int_y^1 \frac{dx}{x} f_{i/N_1}\left(x,\mu_f\right) f_{j/N_2} \left(\frac{y}{x}, \mu_f\right) \equiv f_{i/N_1}(y)\otimes f_{j/N_2}(y)\, .
\ee
The functions $C_{ij}$ in Eq.~(\ref{eq:dMdcos}) are the hard-scattering kernels, which are related to the partonic cross sections and can be calculated in perturbation theory. In order not to clutter the notation, we do not indicate explicitly the fact that the hard-scattering kernels depend on the top-squark masses $m_{\st}$ and $m_{\tilde{t}_2}$, 
the mass $m_{\tilde{q}}$ of the first two generations of squarks and of the sbottoms (which we assume to be all degenerate), the top-quark mass $m_t$, the gluino mass $m_{\tilde{g}}$, and the $\tilde{t}_1$-$\tilde{t}_2$ mixing angle $\alpha$. The expansion of the $C_{ij}$ functions in powers of $\alpha_s$ has the generic form
\be
C_{ij} = \alpha_s^2 \left[ C_{ij}^{(0)} + \frac{\alpha_s}{4 \pi} C_{ij}^{(1)} + \left(\frac{\alpha_s}{4 \pi} \right)^2 C_{ij}^{(2)} + {\mathcal O}(\alpha_s^3)\right] \, .
\label{eq:hardsc}
\ee
Only the quark annihilation and gluon fusion channels contribute to $C_{ij}$ at lowest order in perturbation theory; in particular
\begin{align}
C_{q \bar{q}}^{(0)} &= \delta(1-z) \frac{C_F}{N} \left( \frac{t_1 u_1}{M^4} -\frac{ m_{\st}}{M^2}\right) ,\nonumber \\
C_{gg}^{(0)} &= \delta(1-z) \frac{1}{(N^2-1)} \left(C_F \frac{M^4}{t_1 u_1} -C_A\right) \left( \frac{t_1 u_1}{M^4} -\frac{2 m_{\st}^2}{M^2} +\frac{2 m_{\st}^4}{t_1 u_1}\right) , 
\end{align}
where $N=3$, and the Mandelstam invariants $t_1$ and $u_1$ can be written in terms of $s$ and $\theta$ as
\be
t_1 = -\frac{M^2}{2} \left(1 - \beta_{\st} \cos{\theta} \right) , \qquad u_1 = -\frac{M^2}{2} \left(1 + \beta_{\st} \cos{\theta} \right) . 
\ee
In order to calculate higher-order corrections to $C_{q \bar{q}}$ and $C_{gg}$ one needs to consider virtual and real-emission corrections to the Born approximation. Starting at order $\alpha_s$, new production channels, such as $q g \to \st \st^* q$, open up. When working in the threshold limit
$z \to 1$, the calculations are simplified by the fact that there is no phase-space available for the emission of additional (hard) partons in the final state. 
Consequently, both the hard-gluon emission and the additional production channels are suppressed by powers of $(1-z)$ and can be safely neglected while deriving results within the partonic-threshold limit.
By neglecting power-suppressed terms in the integrand, Eq.~(\ref{eq:dMdcos}) can be rewritten as
\begin{align}
\frac{d^2 \sigma}{d M d \cos{\theta}} &= \frac{\pi \beta_{\st}}{S M} \int_\tau^1 \frac{dz}{z} \Bigl[ \ff_{gg}\left(\frac{\tau}{z},\mu_f \right)
C_{gg}\left(z,M, \cos{\theta},\mu_f \right) \nonumber \\
&  + \ff_{q \bar{q}} \left(\frac{\tau}{z},\mu_f \right) C_{q \bar{q}}\left(z,M, \cos{\theta},\mu_f \right) + 
\ff_{\bar{q} q} \left(\frac{\tau}{z},\mu_f \right) C_{q\bar{q}}\left(z,M,-\cos{\theta},\mu_f \right)  \Bigr] \, .
\label{eq:abx}
\end{align}
In Eq.~(\ref{eq:abx})  the quark channel luminosities $\ff_{q \bar{q}}$ and $\ff_{\bar{q} q}$ are understood to be summed over all light quark flavors. The two terms in the second line of  Eq.~(\ref{eq:abx}) differ in the fact that in the first term the quark (antiquark) comes from the hadron $N_1$ ($N_2$) in Eq.~(\ref{eq:hadproc}), while in the second term the quark (antiquark) comes from the hadron $N_2$ ($N_1$), respectively.
The total cross section can be obtained by integrating over $\cos \theta$ in the range $[-1,1]$ and over $M$ in the range
$[2 m_{\st}, \sqrt{S}]$.

In the soft-gluon emission limit $z \to 1$, the hard-scattering kernels $C_{ij}$ factor into a product of hard and soft functions:
\be
C_{ij}(z,M,\cos{\theta},\mu_f) = \Tr\left[\bm{H}_{ij}(M,\cos{\theta},\mu_f) 
\bm{S}_{ij}(\sqrt{s}(1-z),M,\cos{\theta},\mu_f)\right] + 
{\mathcal O}(1-z) \, . \label{eq:Fact}
\ee
Here and in what follows we employ boldface fonts to indicate matrices in color space, such as the hard functions $\bm{H}_{ij}$ and the soft functions $\bm{S}_{ij}$.
For simplicity we drop the  the top mass and the SUSY parameters from the list of arguments of the hard functions $\bm{H}_{ij}$,  as well as the stop mass from the arguments of the soft functions $\bm{S}_{ij}$. Throughout this paper, we
work in the $s$-channel singlet-octet basis
\begin{gather}
  \big( c^{q\bar{q}}_1 \big)_{\{a\}} = \delta_{a_1a_2} \delta_{a_3a_4}
  \, , \qquad \big( c^{q\bar{q}}_2 \big)_{\{a\}} = t^c_{a_2a_1}
  t^c_{a_3a_4} \, , \nonumber
  \\
  \big( c^{gg}_1 \big)_{\{a\}} = \delta^{a_1a_2} \delta_{a_3a_4} \, ,
  \qquad \big( c^{gg}_2 \big)_{\{a\}} = if^{a_1a_2c} \, t^c_{a_3a_4}
  \, , \qquad \big( c^{gg}_3 \big)_{\{a\}} = d^{a_1a_2c} \,
  t^c_{a_3a_4} \, ,
    \label{eq:colorstructures}
\end{gather}
where $a_i$ represent the color index of the particle with momentum $p_i$.
We view these structures as basis vectors $\ket{c_I}$ in the space of
color-singlet amplitudes. Inner products in this space are defined
through a summation over color indices as
\begin{align}
  \label{eq:innerproducts}
  \Braket{c_I|c_J}= \sum_{\{a\}}\big(c_I\big)^*_{a_1 a_2 a_3 a_4}
  \big(c_J\big)_{a_1 a_2 a_3 a_4} \,.
\end{align}
This inner product is proportional but not equal to $\delta_{IJ}$, so
the basis vectors are orthogonal but not orthonormal. 

A factorization formula analogous to  Eq.~(\ref{eq:Fact}) for the top-quark pair production was derived employing SCET and heavy-quark effective theory in \cite{Ahrens:2010zv}. A completely analogous procedure can be followed in order to derive Eq.~(\ref{eq:Fact}), which is valid in the case of top-squark pair production. 

The hard functions are obtained from the virtual corrections and are ordinary functions of their arguments. The soft functions arise from the real emission of soft gluons and contain distributions which are singular in the $z \to 1$ limit.
Contributions of order $\alpha_s^n$ to the soft functions include terms proportional to the plus distributions
\be\label{eq:plusdistrib}
P_m(z) = \left[\frac{\ln^m(1-z)}{1-z}\right]_+ \, ; \quad m = 0,\dots, 2n -1 \, , 
\ee
as well as terms proportional to $\delta(1-z)$.
In particular, the NLO and NNLO hard-scattering kernels in Eq.~(\ref{eq:hardsc}) have the structure
\begin{align}\label{Cpim}
C^{(1)}_{ij} &= \sum_{m=0}^1 D^{(1)}_{m,ij} P_m(z) + Q^{(1)}_{0,ij}
 \delta(1-z) + R^{(1)}_{ij}(z) \, , \nonumber \\[-2mm]
C^{(2)}_{ij} &= \sum_{m=0}^3 D^{(2)}_{m,ij} P_m(z)
+ Q^{(2)}_{0,ij} \delta(1-z) +   R^{(2)}_{ij}(z) \, . 
\end{align}
The functions $D^{(k)}_{m,ij}$, $Q^{(k)}_{0,ij}$, and $R^{(k)}_{ij}$ depend also on $\cos{\theta}$, $M$, $\mu_f$ and on the heavy-particles masses.
The coefficients $D^{(1)}_{m,ij}$, $Q^{(1)}_{0,ij}$, and $R^{(1)}_{ij}$ can in principle be obtained from results present in the literature. 
One of the main results of this paper is the calculation of the coefficients $D^{(2)}_{m,ij}$ (with $m=0,\dots,3$) both in the quark annihilation and gluon fusion 
channels. We can also evaluate all of the scale-dependent terms in $Q^{(2)}_{0,ij}$ 
in both channels, but due to the ambiguity on the choice of the normalization scale in the argument of these logarithms we drop part of these terms in the numerical implementation of our formulas. We will return to this issue below.

\subsection{1PI kinematics}

The 1PI kinematics approach allows one to describe observables in which a single particle, rather than a pair, is considered. One can then write the top-squark rapidity ($y$) and transverse-momentum ($p_T$) distribution as 
\be
\frac{d^2 \sigma}{d p_T d y} = \frac{2 \pi p_T}{S} \sum_{ij} \int_{x_{1}^{{\rm min}}}^1 \frac{d x_1}{x_1}  \int_{x_{2}^{{\rm min}}}^1 \frac{d x_2}{x_2}
f_{i/N_1}(x_1,\mu_f) f_{j/N_2}(x_2,\mu_f)\,C_{{\rm 1PI}, ij}\left(s_4,s,t_1,u_1,\mu_f\right)  \, ;
\label{eq:1PIdiff}
\ee
once again we will drop the subscript 1PI in most of our equations below. The expansion of the 1PI hard-scattering kernels $C_{{\rm 1PI}}$ in powers of $\alpha_s$ has the same  structure shown in Eq.~(\ref{eq:hardsc}) for the PIM case. Obviously, also in this case only the $q \bar{q}$ and $gg$ channels give non-vanishing contributions at lowest order in $\alpha_s$. The hadronic Mandelstam variables $T_1$ and $U_1$ are related to the stop rapidity and transverse momentum through the relations
\be
T_1 = -\sqrt{S} m_\perp e^{-y} \, , \qquad U_1 = -\sqrt{S} m_\perp e^y \, ,
\ee
where $m_\perp = \sqrt{p_T^2 + m_{\st}^2}$. Therefore, the variables $s,s_4,t_1$, $u_1$, which are arguments of the 1PI hard functions can be expressed in terms of $p_T,y,x_1$, $x_2$. The lower integration limits in Eq.~(\ref{eq:1PIdiff}) are
\be
x_1^{{\rm min}} = - \frac{U_1}{S+T_1} \, , \qquad x_2^{{\rm min}} = - \frac{x_1 T_1}{x_1 S + U_1} \, .
\ee
In order to obtain the total cross section, it is necessary to integrate the double-differential distribution with respect to the top-squark rapidity and transverse momentum over the range
\be
0 \le |y| \le \frac{1}{2} \ln{\frac{1 + \sqrt{1-4 m_\perp^2/S}}{1- \sqrt{1-4 m_\perp^2/S}}} \, , \qquad 0 \le p_T \le \sqrt{\frac{S}{4} - m_{\st}^2} \, .
\ee

In the case of 1PI kinematics, the hard-scattering kernels in the soft-emission limit  $s_4 \to 0$ factor into a product of hard and soft functions, in analogy with Eq.~(\ref{eq:Fact}):
\be\label{soft1PI}
C_{ij}(s_4,s',t'_1,u'_1,\mu) = \Tr \left[\bm{H}_{ij}(s',t'_1,u'_1,\mu) 
\bm{S}_{ij}(s_4,s',t'_1,u'_1,\mu) \right]  + {\mathcal O}(s_4) \, .
\ee
As emphasized in \cite{Ahrens:2011mw}, the Mandelstam invariants $s', t'_1$, $u'_1$ can differ from $s, t_1$, $u_1$ by power corrections proportional to $s_4$. For example, explicit results for the hard and soft functions can be rewritten by employing either the relation $s'+t'_1+u'_1 = 0$ or $s'+t'_1+u'_1 = s_4$.
The difference between the two choices is due to terms suppressed by positive powers of $s_4$.
A detailed description of the way in which we deal with this ambiguity can be found in Section~4 of \cite{Ahrens:2011mw}.

As in the case of PIM kinematics, the hard and soft function are matrices in color space originating from virtual and soft-emission corrections, respectively. 
The 1PI hard functions are identical to the ones encountered in the case of PIM kinematics, provided that the variables $s,t_1$, and $u_1$ are written in terms of $M$ and $\cos\theta$. The soft functions are different in the PIM and 1PI schemes, but in both cases they are identical to the ones employed in the calculation of the top-quark pair production cross sections in \cite{Ahrens:2010zv,Ahrens:2011mw}. 

The 1PI soft functions at order $\alpha_s^n$ depend on the associated plus distributions 
\be\label{eq:associatedplus}
\bar P_m(s_4) = \left[ \frac{\ln^m(s_4/m_{\st}^2)}{s_4}\right]_+ 
= \frac{1}{m_{\st}^2}\,P_m\bigg(1-\frac{s_4}{m_{\st}^2}\bigg)
\, ; \qquad m = 0, \dots, 2 n-1 \, .
\ee
It follows that
\be
\int_0^{s_4^{{\rm max}}} d s_4 \left[\frac{\ln^m(s_4/m_{\st}^2)}{s_4}\right]_+ g(s_4) = \int_0^{s_4^{{\rm max}}} d s_4 \frac{\ln^m(s_4/m_{\st}^2)}{s_4} \left[ g(s_4) - g(0) \right] + \frac{g(0)}{m+1} \, \ln^{m+1}{\frac{s^{{\rm max}}_4}{m_{\st}^2}} \, .
\ee
where $g$ is a smooth test function. 
At NLO and NNLO, the hard-scattering kernel in 1PI kinematics have the structure 
\begin{align}\label{C1pi}
C^{(1)}_{ij} &= \sum_{m=0,1} D^{(1)}_{m,ij} \bar P_m(s_4) + Q^{(1)}_{0,ij}
 \delta(s_4) + R^{(1)}_{ij}(s_4) \, , \nonumber \\[-2mm]
C^{(2)}_{ij}  &= \sum_{m=0}^3 D^{(2)}_{m,ij} \bar P_m(s_4) + Q^{(2)}_{0,ij}
 \delta(s_4) +   R^{(2)}_{ij}(s_4) \, . 
\end{align}
As in the PIM case,  the NLO coefficients $D^{(1)}_{m,ij}$, $Q^{(1)}_{0,ij}$, and $R^{(1)}_{ij}$ can be in principle obtained from 
the literature. In this work we are able to derive exact expressions for the NNLO coefficients $D^{(2)}_{m,ij}$ (with $m=0,\dots,3$) 
and the scale-dependent terms in the coefficients $Q^{(2)}_{0,ij}$.

\section{The hard and soft functions at NLO}
\label{sec:Matching}

In this section we describe the calculation of the soft and hard matrices up to NLO in
perturbation theory.

\subsection{Hard functions}

As it was shown in detail in \cite{Ahrens:2010zv}, the hard functions are defined in terms of the products of Wilson coefficients with their complex conjugates. In order to obtain the Wilson coefficients, one matches renormalized Green's functions in SUSY-QCD with those in SCET. The matching procedure can be carried out by choosing arbitrary external states and infrared (IR) regulators.
The simplest and most common procedure consists in employing on-shell states for the partonic processes 
$(q \bar{q}, gg) \to \tilde{t}_1 \tilde{t}^*_1$ and in using dimensional regularization to regulate both IR and ultraviolet (UV) divergences. With this choice, the SCET loop diagrams vanish because they are scaleless. Consequently, the effective theory matrix elements are equal to their tree-level expressions multiplied by a UV renormalization matrix $\bm{Z}$.
The SUSY-QCD matrix elements are instead the virtual corrections to the scattering amplitudes of the partonic processes
$(q \bar{q}, gg) \to \tilde{t}_1 \tilde{t}^*_1$.

Rather than implementing the matching condition for the Wilson coefficients, we directly move to the calculation of the  
the matrix elements $H_{ij}^{IJ}$ of the hard functions projected onto a certain basis $\{\Ket{c_I}\}$ of color structures.
These matrix elements, and not the Wilson coefficients, are needed in order to obtain the approximate NNLO formulas
which represent the main result of the present work. Throughout this paper, we use the $s$-channel singlet-octet basis defined in Eq.~(\ref{eq:colorstructures}). In order to calculate the hard-function matrix elements, we use the fact 
that the renormalized hard functions can be directly obtained from the corresponding (squared) on-shell QCD scattering amplitudes. The infrared poles can be removed from the QCD amplitudes by employing the prescription of \cite{Becher:2009cu,Becher:2009qa}
\begin{align}
  \label{eq:renM}
  \Ket{\mathcal{M}_{ij}^{\text{ren}}} \equiv \lim_{\epsilon \to 0} \bm{Z}_{ij}^{-1}(\epsilon)
  \Ket{\mathcal{M}_{ij}(\epsilon)} = 4\pi\alpha_s \left[ \Ket{\mathcal{M}_{ij}^{\text{ren}\,(0)}} +
    \frac{\alpha_s}{4\pi} \Ket{\mathcal{M}_{ij}^{\text{ren}\,(1)}} + \dots \right] \, ,
\end{align}
where the indices $ij$ are not summed over.
Here $\Ket{\mathcal{M}_{ij}(\epsilon)}$ are the dimensionally regularized (and UV renormalized) scattering amplitudes, whose IR poles are removed by the factors $\bm{Z}_{ij}^{-1}(\epsilon)$. 
The resulting finite amplitudes are expressed in terms of $\alpha_s$ with $n_l=5$ active flavors. The explicit form of the $\bm{Z}_{ij}$ matrices can be obtained by employing the results derived in \cite{Becher:2009kw,Ferroglia:2009ep,Ferroglia:2009ii}.

The perturbative expansion of the renormalized hard functions is defined as 
\begin{align}
  \bm{H}_{ij} = \alpha_s^2 \, \frac{1}{d_R} \left( \bm{H}_{ij}^{(0)} + \frac{\alpha_s}{4\pi}
    \bm{H}_{ij}^{(1)} + \dots \right) \, ,
\end{align}
where $d_R = N$ in the quark annihilation channel and $d_R = N^2-1$ in the gluon fusion channel.
The desired matrix elements can then be written in terms of the renormalized QCD amplitudes and the color basis vectors $\Ket{c_I}$ as
\begin{align}
  \label{eq:hform}
  H_{ij}^{(0)\,IJ} &= \frac{1}{4} \, \frac{1}{\braket{c_I|c_I}\braket{c_J|c_J}} \Braket{c_I |
    \mathcal{M}_{ij}^{\text{ren}\,(0)}} \Braket{\mathcal{M}_{ij}^{\text{ren}\,(0)} | c_J} \, ,
  \nonumber
  \\
  H_{ij}^{(1)\,IJ} &= \frac{1}{4} \, \frac{1}{\braket{c_I|c_I}\braket{c_J|c_J}} \bigg[
  \Braket{c_I | \mathcal{M}_{ij}^{\text{ren}\,(0)}} \Braket{\mathcal{M}_{ij}^{\text{ren}\,(1)} |
    c_J} + \Braket{c_I | \mathcal{M}_{ij}^{\text{ren}\,(1)}}
  \Braket{\mathcal{M}_{ij}^{\text{ren}\,(0)} | c_J} \bigg] \, .
\end{align}
The leading-order result for the $q\bar q$ channel follows from a simple calculation. In matrix notation, it reads 
\begin{align}
  \bm{H}_{q\bar{q}}^{(0)} =
  \begin{pmatrix}
    0~ & 0
    \\
    0~ & 1
  \end{pmatrix}
  \frac{2}{t_1+u_1}
  \Bigg( \frac{t_1 u_1}{t_1 + u_1} + m_{\tilde t_1}^2 \Bigg) \, ,
\end{align}
while that for the $gg$ channel is  
\begin{align}
  \bm{H}_{gg}^{(0)}  =
  \begin{pmatrix}
    \frac{1}{N^2} & \frac{1}{N}\,\frac{t_1-u_1}{M^2} & \frac{1}{N}
    \\
    \frac{1}{N}\,\frac{t_1-u_1}{M^2} & \frac{(t_1-u_1)^2}{M^4} & \frac{t_1-u_1}{M^2}
    \\
    \frac{1}{N} & \frac{t_1-u_1}{M^2} & 1
  \end{pmatrix}
  \frac{m_{\tilde t_1}^4}{2t_1u_1} \Bigg(\! -\frac{2M^2}{m_{\tilde t_1}^2}+\frac{t_1 u_1}{m_{\tilde t_1}^4} + 4 
  +2 \frac{t_1^2+u_1^2}{t_1u_1} \Bigg) \, .
\end{align}

In order to calculate the NLO hard matrices $\bm{H}_{ij}^{(1)}$, one needs to evaluate the one-loop corrections to the
partonic scattering amplitudes, by keeping the various color components separate. Although results
for the corresponding one-loop diagrams interfered with the Born-level amplitudes are well known \cite{Beenakker:1997ut}, their decomposition into the color basis is not available in the literature, and we therefore had to calculate it from scratch. For this purpose, we have used in-house routines written in the computer
algebra system FORM \cite{Vermaseren:2000nd} in combination with {\tt Reduze} \cite{Studerus:2009ye, vonManteuffel:2012yz}. 
The results of the calculation are expressed in terms of Passarino-Veltman functions \cite{Passarino:1978jh}. After UV renormalization, we have derived analytic expressions for the IR poles, and we have  evaluated the finite parts of the amplitudes numerically by employing the programs described in 
\cite{Hahn:1998yk, Ellis:2007qk, vanHameren:2010cp}.
We have checked our results in several ways. First, by applying the IR renormalization factors $\bm{Z}_{ij}$ of Eq.~(\ref{eq:renM}), we find that the IR poles 
cancel exactly. 
Second, we have checked that by multiplying the one-loop hard functions with the corresponding 
tree-level soft functions and by subsequently taking the trace of the resulting color-space matrices,
we reproduce the numerical results for the NLO virtual corrections which can be extracted from the code
{\tt Prospino} \cite{Beenakker:1996ed}.

The output of our FORM codes are too long to be presented here in explicit form. The explicit expressions for the NLO hard functions are coded in a {\tt Mathematica} and a Fortran program which are included in the arXiv version of this work. In particular, these programs allow one to evaluate numerically the LO and NLO hard functions for an arbitrary choice of the input parameters.

\subsection{Soft functions}

The soft functions are vacuum expectation values of soft Wilson-loop operators. These functions are not sensitive to the spin of the particles involved, but they depend on the color structure of the underlying partonic subprocesses. Consequently, the soft functions needed for the calculation of the top-squark pair-production cross section are precisely the same functions employed in the calculation of the cross section for top-quark pair production. The calculation of the PIM soft functions at NLO was  described in detail in \cite{Ahrens:2010zv}, while
the analogous calculation of the 1PI soft functions was carried out in \cite{Ahrens:2011mw}. For the convenience of the reader, the explicit results for the soft functions obtained in those two papers are collected in Appendix~\ref{SFA}. 

\section{Structure of the hard scattering kernels}
\label{sec:RGEeq}

\subsection{Preliminaries}

Since the soft functions in (\ref{eq:Fact}) and (\ref{soft1PI}) depend on plus distributions, it is more convenient to work with the Laplace-transformed functions
\begin{equation}
\tilde{\bm{s}}_{ij}(L,M,\cos\theta,\mu) \equiv \frac{1}{\sqrt{s}} \int_0^\infty d \omega 
\exp{\left( -\frac{\omega}{e^{\gamma_E} \mu e^{L/2}}\right)} \bm{S}_{ij}(\omega,M,\cos\theta,\mu) \, , 
\label{eq:SoftLaplace}
\end{equation}
in PIM and similarly in the case of 1PI kinematics. It was shown in \cite{Becher:2006nr, Becher:2006mr, Becher:2007ty, Ahrens:2009uz}
that the leading singular terms in the hard-scattering kernels can be generated by replacing the Laplace variable $L$ with a derivative $\partial_\eta$ with respect to an auxiliary variable $\eta$, which is later set to 0. To this end, one defines
Laplace-transformed hard-scattering kernels in PIM kinematics as
\begin{align}
\tilde{c}_{ij}( \partial_\eta, M,\cos{\theta},\mu ) &= \Tr[\bm{H}_{ij}(M,\cos\theta,\mu) \tilde{\bm{s}}_{ij}(\partial_\eta, M,\cos\theta,\mu)] \, .
\end{align}
For 1PI kinematics one replaces the arguments $M$ and $\cos\theta$ with $s'$, $t'_1$, and $u'_1$. 
The hard-scattering kernel in momentum space can then be recovered through the relations \cite{Ahrens:2009uz,Ahrens:2010zv, Ahrens:2011mw}
\begin{align}
C_{{\rm PIM},ij}(z, M,\cos{\theta},\mu) &= \tilde{c}_{ij}(\partial_\eta, M,\cos{\theta},\mu) \left. \left(\frac{M}{\mu}\right)^{2\eta} \frac{e^{-2 \gamma_E \eta}}{\Gamma\left(2 \eta \right)} \frac{z^{-\eta}}{(1-z)^{1- 2 \eta}} \right|_{\eta = 0} , \nonumber \\
C_{{\rm 1PI},ij}\left(s_4,s',t'_1,u'_1,\mu \right) &= \tilde{c}_{ij}(\partial_\eta, s', t'_1, u'_1,\mu ) \left. \frac{e^{-2 \gamma_E \eta}}{\Gamma\left(2 \eta \right)} \frac{1}{s_4} \left(\frac{s_4}{\sqrt{m_{\st}^2 +s_4} \mu} \right)^{2 \eta}  \right|_{\eta =0} \, .
\end{align}
In order to evaluate the above formulas, one needs to employ analytic continuation to regulate the divergences in $z \to 1$ or $s_4 \to 0$, to take the derivatives with respect to $\eta$, and finally to take the limit $\eta \to 0$. It is possible to show that the result of this procedure can equivalently be obtained by implementing a series of replacement rules on the functions $\tilde{c}_{ij}\left(L,\dots \right)$ considered as polynomials in a new variable $L$. In 
PIM kinematics, one must substitute
\begin{align}
  1 &\to \delta(1-z) \, , \nonumber
  \\
  L &\to 2 P_0(z) + \delta(1-z) L_M, \nonumber
  \\
  L^2 &\to 8 P_1(z) +4 P_0(z) L_M+ \delta(1-z) \left(L^2_M -\frac{2}{3} \pi^2\right) -4 \frac{\ln z}{1-z} , \nonumber
  \\
  L^3 &\to 24 P_2(z) + 24 P_1(z) L_M + P_0(z) \left( 6 L_M^2  -4 \pi^2 \right) + \delta(1 - z) \Bigl( L_M^3
   - 2 \pi^2 L_M
   +16 \zeta_3\Bigr)
        \nonumber \\
        &
  + \frac{6}{1-z} \Bigl[ \ln^2 z - 4 \ln z \ln(1 - z)
            -  2 \ln z L_M
            \Bigr]\, , \nonumber
  \\
  L^4 &\to 64 P_3(z)  + 96 P_2(z) L_M +P_1(z) \left( 48 L^2_M 
-32 \pi^2 \right) + P_0(z) \Bigl(8 L_M^3
-16 \pi^2 L_M  +128 \zeta_3\Bigr) 
\nonumber \\ &+ \delta(1-z) \left( L_M^4
 - 4 \pi^2 L_M^2 + 64 \zeta_3 L_M
 + \frac{4}{15} \pi^4\right) +\frac{8}{1-z} \Bigl[ 2 \pi^2 \ln{z}
           - 12\ln{z} \ln^2 {(1-z)}
            \nonumber \\ & 
           - 12\ln{z} \ln{(1-z)}  L_M
           - 3\ln{z} L_M^2
           + 6\ln^2{z} \ln{(1-z)}
           + 3 \ln^2{z}  L_M
           - \ln^3{z} \Bigr] \, ,
\end{align}
where $L_M = \ln{(M^2/\mu^2)}$, and the plus distributions $P_m(z)$ have been defined in Eq.~(\ref{eq:plusdistrib}).
In 1PI kinematics, one must substitute instead
\begin{align}
  1 &\longrightarrow \delta(s_4) \, , \nonumber
  \\
  L &\longrightarrow 2 \bar P_0(s_4) - \delta(s_4) \, L_m \, , \nonumber
  \\
  L^2 &\longrightarrow 8 \bar P_1(s_4) - 4 L_m \bar P_0(s_4) + \delta(s_4) \left( L_m^2 -
    \frac{2\pi^2}{3} \right) - \frac{4L_4}{s_4} \, , \nonumber
  \\
  L^3 &\longrightarrow 24 \bar P_2(s_4) - 24L_m \bar P_1(s_4) + \left( 6L_m^2 - 4\pi^2 \right)
  \bar P_0(s_4) + \delta(s_4) \Big( \! -L_m^3 + 2\pi^2 L_m + 16\zeta_3 \Big) \nonumber
  \\
  &\quad - \frac{6L_4}{s_4} \left( 2 L_s -L_4\right) ,
  \nonumber
  \\
  L^4 &\longrightarrow 64 \bar P_3(s_4) - 96L_m \bar P_2(s_4) + \left( 48L_m^2 - 32\pi^2 \right)
  \bar P_1(s_4) + \Big( \! -8L_m^3 + 16\pi^2 L_m + 128\zeta_3 \Big) \bar P_0(s_4) \nonumber
  \\
  &\quad + \delta(s_4) \left( L_m^4 - 4\pi^2 L_m^2 - 64\zeta_3 L_m + \frac{4\pi^4}{15}
  \right) 
   - \frac{8L_4}{s_4} \Big( L_4^2 - 3 L_4 L_s + 3
    L_s^2 - 2\pi^2 \Big) ,
\label{eq:conversion}
\end{align}
where $L_m=\ln(\mu^2/m_{\st}^2)$, $L_4=\ln(1+s_4/m_{\st}^2)$, $L_s = \ln [s_4^2/(m_{\st}^2 \mu^2)]$, and the associated plus distributions $\bar P_m(s_4)$ have been defined in Eq.~(\ref{eq:associatedplus}).

Since the hard and soft functions are known up to NLO, is easy to determine the NLO coefficient in the expansion of $\tilde{c}$ in powers of $\alpha_s$:
\be
\tilde{c}_{ij} = \alpha_s^2\left[ \tilde{c}_{ij}^{(0)} + \frac{\alpha_s}{4 \pi} \tilde{c}_{ij}^{(1)} + \left(\frac{\alpha_s}{4 \pi}\right)^2 \tilde{c}_{ij}^{(2)} + {\mathcal O}(\alpha_s^3) \right] .
\ee
It is important to observe that the trace of the product of the LO hard function and NLO soft function contains the dependence of $\tilde{c}_{ij}^{(1)}$ on $L$, and therefore it gives rise to the plus distributions.
In order to obtain the complete coefficient $\tilde{c}_{ij}^{(2)}$ one needs to know the hard and soft functions up to NNLO:
\be
\tilde{c}_{ij}^{(2)} = \frac{1}{d_R} \left\{\Tr\left[\bm{H}_{ij}^{(1)} \tilde{\bm{s}}_{ij}^{(1)}\right] + \Tr\left[\bm{H}_{ij}^{(0)} \tilde{\bm{s}}_{ij}^{(2)}\right] + \Tr\left[\bm{H}_{ij}^{(2)} \tilde{\bm{s}}_{ij}^{(0)}\right]\right\}
= \sum_{k=0}^4 c^{(2)}_{k,ij} L^k\, . \label{eq:NNLOapp}
\ee
The NLO hard and soft functions in this equation are know exactly. As explained in detail in \cite{Ahrens:2010zv,Ahrens:2011mw}, the scale-dependent part of the NNLO hard function and the $L$ dependent part of the NNLO soft function can  be reconstructed by exploiting the information coming from the RGE satisfied by these functions. To this end, one only needs the one-loop hard and soft matrices calculated in Section~\ref{sec:Matching} and in \cite{Ahrens:2010zv,Ahrens:2011mw}, respectively, as well as the relevant two- and three-loop anomalous dimensions computed in \cite{Ferroglia:2009ep,Ferroglia:2009ii,Korchemskaya:1992je,Moch:2004pa}. After this information on the NNLO hard and soft function has been extracted, one can evaluate the coefficients $c^{(2)}_{k,ij}$ with $k=1,\dots,4$ in Eq.~(\ref{eq:NNLOapp}), as well as the scale-dependent part of $c^{(2)}_{0,ij}$.

\subsection{Approximate NNLO formulas}

\label{sec:AppNNLO}

\begin{table}
\begin{center}
\begin{tabular}{|c|c|c|c|}
\hline $s$ &  $9.0 \times 10^6~\rm{GeV}^2$ &  $m_{\tilde{t}_2}$ & $ 1319.87$~GeV\\ 
\hline $t_1$ &  $-2.94979\times 10^6~\rm{GeV}^2$& $m_{\tilde{q}}$  & $1460.3$~GeV \\ 
\hline  $\mu$ & $1087.17$~GeV  &  $m_t$ & $173.3$~GeV \\ 
\hline $m_{\st}$ & $1087.17$~GeV & $\alpha$ & $68.4^\circ$ \\ 
\hline $m_{\tilde{g}}$ & $1489.98$~GeV  &  &  \\ 
\hline 
\end{tabular}
\end{center}
\caption{Compilation of input parameters employed in the calculation of the hard-scattering coefficients. The angle $\alpha$ is the top-squark mixing angle. \label{tab:inputs}}
\end{table} 

By employing the results described in the previous sections, we are able to obtain approximate NNLO formulas for the hard-scattering kernels in Eqs.~(\ref{eq:Fact}) and (\ref{soft1PI}). 
These formulas include the exact expressions of the coefficients multiplying the plus distributions up to NNLO, both in PIM and in 1PI kinematics. The complete results for these coefficients, written in terms of Passarino-Veltman functions, are too lengthy to be reported here. The values of the coefficients multiplying the various plus distributions and delta functions for arbitrary values of the input parameters can be extracted from the {\tt Mathematica} code mentioned above, which we include in the arXiv submission of this work.\footnote{The {\tt Mathematica} file requires the use of {\tt LoopTools} \cite{Hahn:1998yk}.}
As a reference for the reader, we present here the hard scattering kernels evaluated by setting the  input parameters at the values listed in Table~\ref{tab:inputs}.  The SUSY spectrum which we chose corresponds to the benchmark point {\tt 40.2.5} in \cite{AbdusSalam:2011fc}; we employ this benchmark point for the remainder of this paper. 
By defining $\hat{C}^{(i)} = d_R C^{(i)}$ ($i=0,1,2$), in the quark annihilation channel with PIM kinematics one finds
\begin{align}
\hat{C}^{(0)}_{{\rm PIM},q \bar{q}}(z) &= 0.118673 \, \delta(1-z)\, , \nonumber \\
\hat{C}^{(1)}_{{\rm PIM},q \bar{q}}(z) &=  2.53170 \left[\frac{\ln (1-z)}{1-z}\right]_+ +
1.18594 \left[\frac{1}{1-z}\right]_+ + 0.834825 \,  \delta(1-z)  +\dots \, ,\nonumber \\
\hat{C}^{(2)}_{{\rm PIM},q \bar{q}}(z) &= 27.0048 \left[\frac{\ln^3 (1-z)}{1-z}\right]_+ +18.5403  \left[\frac{\ln^2 (1-z)}{1-z}\right]_+ 
 -56.3923\left[\frac{\ln (1-z)}{1-z}\right]_+ \nonumber \\ 
    & + 62.2067  \left[\frac{1}{1-z}\right]_+ 
 -(29.5324+82.8060) \,  \delta(1-z)  +\dots  \, ,
    \label{eq:PIMrefpointqq}
\end{align}
where the ellipses indicate terms which are subleading (and finite) in the $z \to 1$ limit. In the gluon fusion channel, we find instead
\begin{align}
 \hat{C}^{(0)}_{{\rm PIM},gg}(z) &=  0.348572 \, \delta(1-z)\, , \nonumber \\
 \hat{C}^{(1)}_{{\rm PIM},gg}(z) &=  16.7315  \left[\frac{\ln (1-z)}{1-z}\right]_+ +
13.6194 \left[\frac{1}{1-z}\right]_+ + 9.50848 \,  \delta(1-z)  +\dots \, ,\nonumber \\
\hat{C}^{(2)}_{{\rm PIM},gg}(z) &= 401.555\left[\frac{\ln^3 (1-z)}{1-z}\right]_+ +
852.324 \left[\frac{\ln^2 (1-z)}{1-z}\right]_+ 
-389.724  \left[\frac{\ln (1-z)}{1-z}\right]_+ \nonumber \\ 
    & +\,535.481 \left[\frac{1}{1-z}\right]_+ 
+(81.9942-466.336) \,  \delta(1-z)  +\dots  \, .
    \label{eq:PIMrefpointgg}
\end{align}
In order to completely determine the coefficients multiplying the delta~functions in the NNLO hard-scattering kernels, one would need to know the complete NNLO hard and soft matrices. Since those matrices are at present not fully known at NNLO, we are only able to determine the scale-dependent terms in the delta-function coefficients, because those terms are governed by RGEs. 
 However, since the scale-independent parts are unknown, the coefficients of the  delta~function at NNLO depend  on an arbitrary second scale chosen to normalize the scale-dependent logarithms.  In fact, for any renormalization scale $\mu$ and reference scale $\mu_0$, one can always rewrite
\be
\ln\left(\frac{\mu_0^2}{\mu^2} \right) = \ln\left(\frac{\mu_1^2}{\mu^2} \right) + \ln\left(\frac{\mu_0^2}{\mu_1^2} \right) ,
\ee
where the second term in the r.h.s. can be reabsorbed in the unknown $\mu$-independent piece.
 In Eq.~(\ref{eq:PIMrefpointqq}) and Eq.~(\ref{eq:PIMrefpointgg}), we chose the reference scale equal to the pair invariant mass $M$.  
 Furthermore, in Eq.~(\ref{eq:PIMrefpointqq}) and Eq.~(\ref{eq:PIMrefpointgg}), we decided to separate out the contributions to the delta-function coefficients coming from the NNLO hard functions.
In particular, in the round brackets multiplying the delta~functions, the second number represents the  contributions of the scale-dependent terms in ${\bm H}_{ij}^{(2)}$.
In our numerical analysis in Section~\ref{sec:CS}, we decided to drop the contributions to the NNLO delta-function coefficients arising from the two-loop hard functions. As was observed in \cite{Ahrens:2011mw}, this choice is motivated by the fact that by including these $\mu$-dependent terms one might induce an artificial reduction of the scale dependence, which can  lead to an underestimated scale uncertainty.

Similarly, in 1PI kinematics we find for the quark annihilation channel
\begin{align}
\hat{C}^{(0)}_{{\rm 1PI},q \bar{q}}(s_4) &= 0.118673 \, \delta(s_4)\, , \nonumber \\
\hat{C}^{(1)}_{{\rm 1PI},q \bar{q}}(s_4) &= 2.53170  \left[\frac{\ln (s_4/m_{\st}^2)}{s_4}\right]_+ -
 2.03883  \left[\frac{1}{s_4}\right]_+ + 1.66798 \, \delta(s_4)  +\dots \, ,\nonumber \\
\hat{C}^{(2)}_{{\rm 1PI},q \bar{q}}(s_4) &= 27.0048 \left[\frac{\ln^3 (s_4/m_{\st}^2)}{s_4}\right]_+ 
-84.6523 \left[\frac{\ln^2 (s_4/m_{\st}^2)}{s_4}\right]_+ + 34.0042 \left[\frac{\ln (s_4/m_{\st}^2)}{s_4}\right]_+ \nonumber \\ 
    & + 98.7353  \left[\frac{1}{s_4}\right]_+ 
  -(166.272+82.8060) \, \delta(s_4)  +\dots  \, ,
  \label{eq:1PIrefpointqq}
\end{align}
while for the gluon fusion channel we obtain
\begin{align}
 \hat{C}^{(0)}_{{\rm 1PI},gg}(s_4) &=   0.348572 \, \delta(s_4)\, , \nonumber \\
 \hat{C}^{(1)}_{{\rm 1PI},gg}(s_4) &= 16.7315 \left[\frac{\ln (s_4/m_{\st}^2)}{s_4}\right]_+ 
-7.69240 \left[\frac{1}{s_4}\right]_+ + 7.68986 \, \delta(s_4)  +\dots \, ,\nonumber \\
\hat{C}^{(2)}_{{\rm 1PI},gg}(s_4) &=  401.555 \left[\frac{\ln^3 (s_4/m_{\st}^2)}{s_4}\right]_+ 
-682.128 \left[\frac{\ln^2 (s_4/m_{\st}^2)}{s_4}\right]_+  -512.616 \left[\frac{\ln (s_4/m_{\st}^2)}{s_4}\right]_+ \nonumber \\ 
&
 +  1511.18 \left[\frac{1}{s_4}\right]_+   -(1299.25+466.336) \,   \delta(s_4)  +\dots  \, .
  \label{eq:1PIrefpointgg}
\end{align}
In this context, the ellipses indicate subleading terms in the $s_4 \to 0$ limit. Also, in Eqs.~(\ref{eq:1PIrefpointqq}) and (\ref{eq:1PIrefpointgg}), as in the PIM case, the scale-dependent terms are normalized to the scale $s$, which is equal to  $M^2$ when $s_4=0$. As in the PIM case, also in Eqs.~(\ref{eq:1PIrefpointqq}) and (\ref{eq:1PIrefpointgg}) the second number in the round brackets multiplying the delta functions in the NNLO kernels represents the contribution of the scale-dependent terms arising from the NNLO hard functions.

\section{Total cross section}
\label{sec:CS}

In this section we present a brief numerical study of the top-squark  total pair production cross section at approximate NNLO accuracy. We stress that our approximate NNLO predictions include the full NLO corrections, as well as the part of the NNLO corrections arising from soft gluon emission, obtained by means of the procedure outlined in the previous sections.

Following what was done in the study of the top-quark pair production total cross section, we employ $\PIMs$  and 
$\IPIs$ kinematic schemes described in \cite{Ahrens:2011px}. These schemes also include, on top of contributions which are singular in the soft limit, NNLO terms which are regular in the $z \to 1$ ($\PIMs$) or $s_4\to 0$ ($\IPIs$) limits, and which naturally arise from the SCET formalism. 
These terms, which are part of the functions $R_{ij}^{(2)}$ in Eqs.~(\ref{Cpim}) and (\ref{C1pi}),
 do not represent the complete contribution to the NNLO cross section which is regular in the soft limit, since this quantity can only be obtained with a full calculation of this observable at NNLO accuracy.
However, as shown in \cite{Ahrens:2010zv, Ahrens:2011mw}, the regular terms appearing in the  $\PIMs$ and 
$\IPIs$ kinematic approaches arise from the exact definition of the soft-gluon emission energy, and they improve the agreement between exact and approximate formulas at NLO.

Unless otherwise stated, we present results which are obtained by averaging the ones obtained in the two kinematic schemes that we consider. In analogy with what was done in \cite{Ahrens:2011px} for the top-quark pair production cross section, we adopt a conservative approach and consider the difference between the predictions in the two kinematic schemes as an estimate of the uncertainty associated with the use of approximate NNLO formulas.
This is justified by the fact that the two schemes neglect different power-suppressed terms, which are formally subleading but which can nevertheless have a noticeable numerical impact on the total cross section.   To account for this uncertainty, the scale variation of the total cross section is obtained by setting the renormalization and factorization
scales equal to each other, $\mu_R = \mu_f = \mu$, and by varying this single scale between $m_{\tilde{t}_1}/2 < \mu < 2 m_{\tilde{t}_1}$. We then look at the difference between the largest and smallest values obtained. In summary, the central value and perturbative uncertainties for the combined results at approximate NNLO accuracy are  determined by employing the definitions
\begin{align}\label{scalencertainty}
\sigma &= \frac{1}{2} \left( \sigma_{\mbox{\tiny{PIM}}} +  \sigma_{\mbox{\tiny{1PI}}} \right) ,  \nonumber \\
\Delta \sigma^+_\mu &= \mbox{max} \left\{\sigma_{\mbox{\tiny{PIM}}} + \Delta \sigma^+_{\mbox{\tiny{PIM}}} ,  \sigma_{\mbox{\tiny{1PI}}} + \Delta \sigma^+_{\mbox{\tiny{1PI}}} \right\}  - \sigma\nonumber \, , \\
\Delta \sigma^{-}_\mu & = \mbox{min} \left\{\sigma_{\mbox{\tiny{PIM}}} + \Delta \sigma^-_{\mbox{\tiny{PIM}}} ,  \sigma_{\mbox{\tiny{1PI}}} + \Delta \sigma^-_{\mbox{\tiny{1PI}}} \right\}  - \sigma \, ,
\end{align}
where the subscripts $\mbox{1PI}$ and $\mbox{PIM}$ indicate that the corresponding quantities are evaluated in  $\IPIs$ and $\PIMs$ kinematics, respectively, including the full set of NLO corrections and the contribution of the NNLO terms present in the approximate formulas for that scheme. 

As will be shown later, the total cross section is strongly dependent on the mass of the produced particle, $m_{\tilde{t}_1}$. However, similarly to the slepton-pair production case \cite{Broggio:2011bd}, the dependence of the total cross section on other SUSY parameters is relatively small.
In order to show that this is indeed the case, we fix the value of the stop mass equal to $m_{\st}=1087.17$~GeV, corresponding to the SUSY benchmark point {\tt 40.2.5} of 
\cite{AbdusSalam:2011fc} (which is defined in Table~\ref{tab:inputs}), and we provide predictions for the total cross section for three different sets of the remaining SUSY parameters. The first set of SUSY parameters coincides with  the benchmark point {\tt 40.2.5} itself.  The second set is defined by choosing $m_{\tilde{t}_2}=2639.74$~GeV, $m_{\tilde{q}}=2920.61$~GeV, $m_{\tilde{g}}= 2979.96$~GeV, $\alpha=136.8^{\circ}$. The values above are the double of the corresponding values of the benchmark point {\tt 40.2.5}, and for this reason we label 
this second set of SUSY parameters as ``double''. (The value of $m_t$ is kept fixed at $173.3$~GeV.) 
Similarly, the third set of SUSY parameters (which we refer to as ``half'') is $m_{\tilde{t}_2}=659.93$~GeV, $m_{\tilde{q}}=730.15$~GeV, $m_{\tilde{g}}=744.99$~GeV, $\alpha=34.2^{\circ}$. The values of the total cross sections, with the relative perturbative uncertainty, for the three different SUSY parameters sets discussed above, can be found in Table~\ref{tab:susydep}. The table refers to a collider energy of $14$~TeV and the PDFs employed are from the
MSTW2008 global fit \cite{Martin:2009iq,Martin:2009bu,Martin:2010db}. 
We observe that the numerical values for the cross section are very close to each other, in spite of the fact that the input SUSY parameters are considerably different in the three sets. This is true both at NLO and at approximate NNLO accuracy.

\begin{table}[t]
\centering
\begin{tabular}{|c||c|c|}
\hline
LHC $14$~TeV & \multicolumn{2}{|c|}{MSTW2008} \\
\hline
SUSY point & $m_{\st}$~[GeV] & 1087.17 \\
\hline
\hline
{\tt 40.2.5} & $(\sigma \pm \Delta \sigma_\mu)_{\rm{{\tiny NLO}}}$ [pb] & $ 44.2^{+4.9}_{-6.0} \times 10^{-4}$\\
\hline
``double" & $(\sigma \pm \Delta \sigma_\mu)_{\rm{{\tiny NLO}}}$ [pb]&  $44.5^{+5.1}_{-6.1} \times 10^{-4}$ \\
\hline
``half" &$(\sigma \pm \Delta \sigma_\mu)_{\rm{{\tiny NLO}}}$ [pb] & $42.4^{+4.0}_{-5.4} \times 10^{-4}$ \\
\hline
\hline
{\tt 40.2.5} & $(\sigma \pm \Delta \sigma_\mu)_{\rm{{\tiny approx. NNLO}}}$ [pb] & $44.3^{+1.3}_{-2.2} \times 10^{-4}$\\
\hline
``double" & $(\sigma \pm \Delta \sigma_\mu)_{\rm{{\tiny approx. NNLO}}}$ [pb]& $44.7^{+1.3}_{-2.3} \times 10^{-4}$  \\
\hline
``half" &$(\sigma \pm \Delta \sigma_\mu)_{\rm{{\tiny approx. NNLO}}}$ [pb] & $42.3^{+ 0.6}_{-1.8} \times 10^{-4}$\\
\hline
\end{tabular}
\caption{Stop-pair production cross sections at the LHC with $\sqrt{S}=14$~TeV for three different sets of the SUSY parameters described in the text. The stop mass is fixed to $m_{\st}=1087.17$~GeV. The numbers are obtained by using MSTW2008 PDFs
\cite{Martin:2009iq,Martin:2009bu,Martin:2010db}. 
\label{tab:susydep}}
\end{table}

In all of the plots and tables discussed below, the SUSY parameters  are set to the values corresponding to the benchmark point {\tt 40.2.5} in Table~\ref{tab:inputs}. This applies also to the lightest stop mass, with the exception of the figures in which we plot the cross section as a function of the stop mass or when a different choice of the stop mass is explicitly indicated.

All of the numbers and plots are obtained by means of an in-house Fortran code, in which the approximate NNLO formulas are implemented.
The NLO calculations, which are one of the elements needed to obtain predictions at approximate NNLO accuracy, have been carried out by modifying the public version of {\tt Prospino} \cite{Beenakker:1996ed}.

\begin{figure}[t]
\begin{center}
\begin{tabular}{cc}
\includegraphics[width=0.48\textwidth]{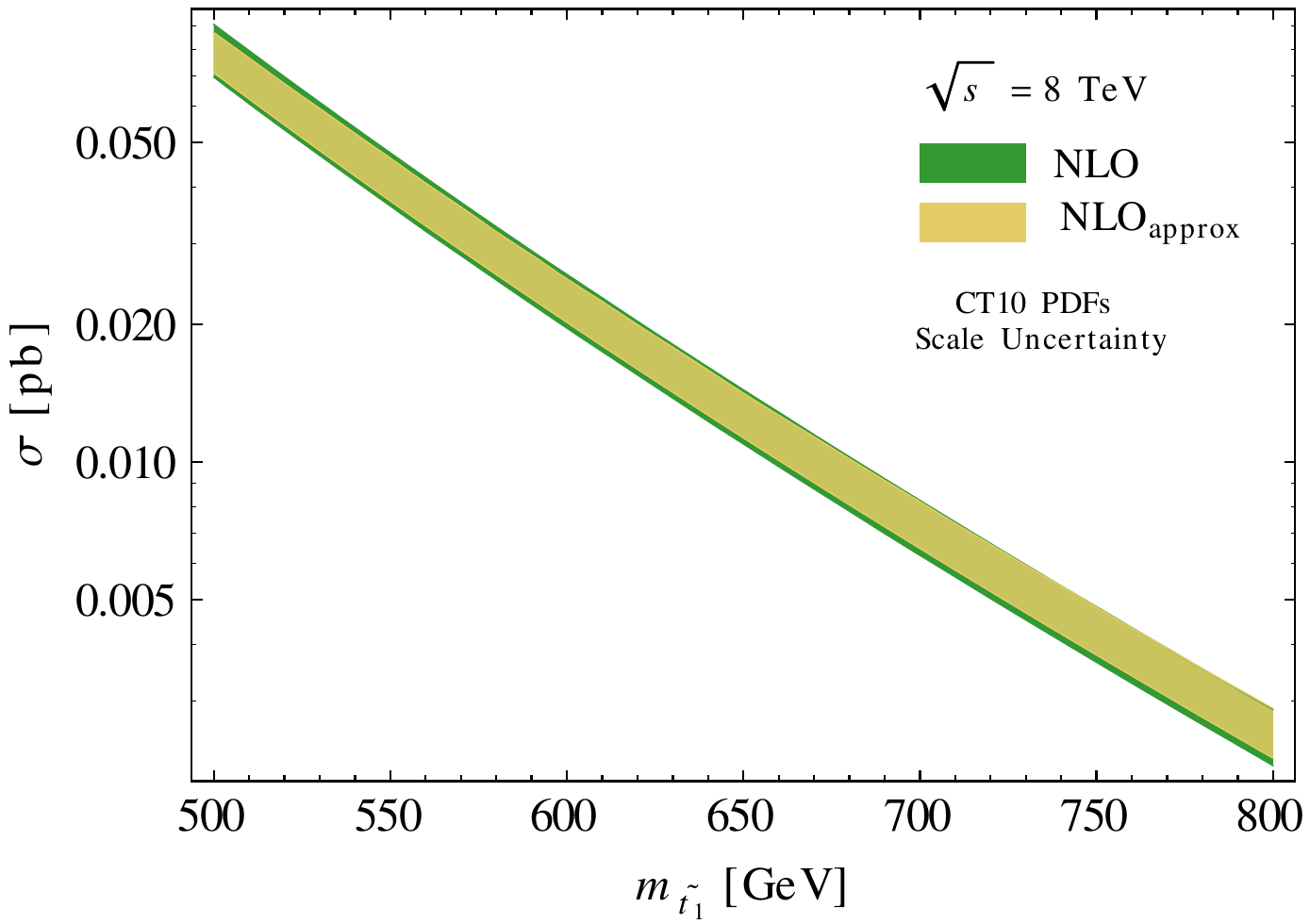} &  \includegraphics[width=0.48\textwidth]{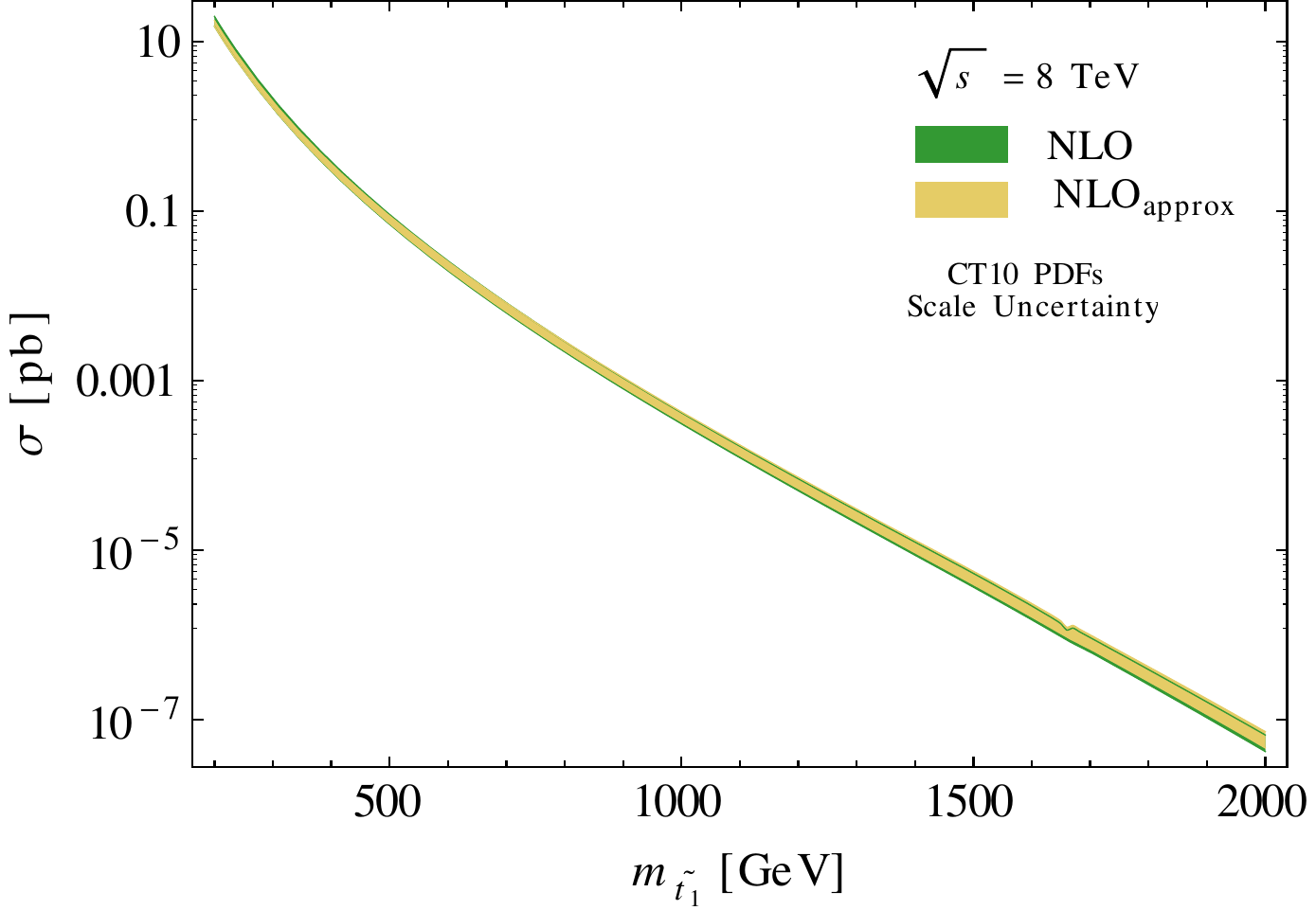}\\ 
\includegraphics[width=0.48\textwidth]{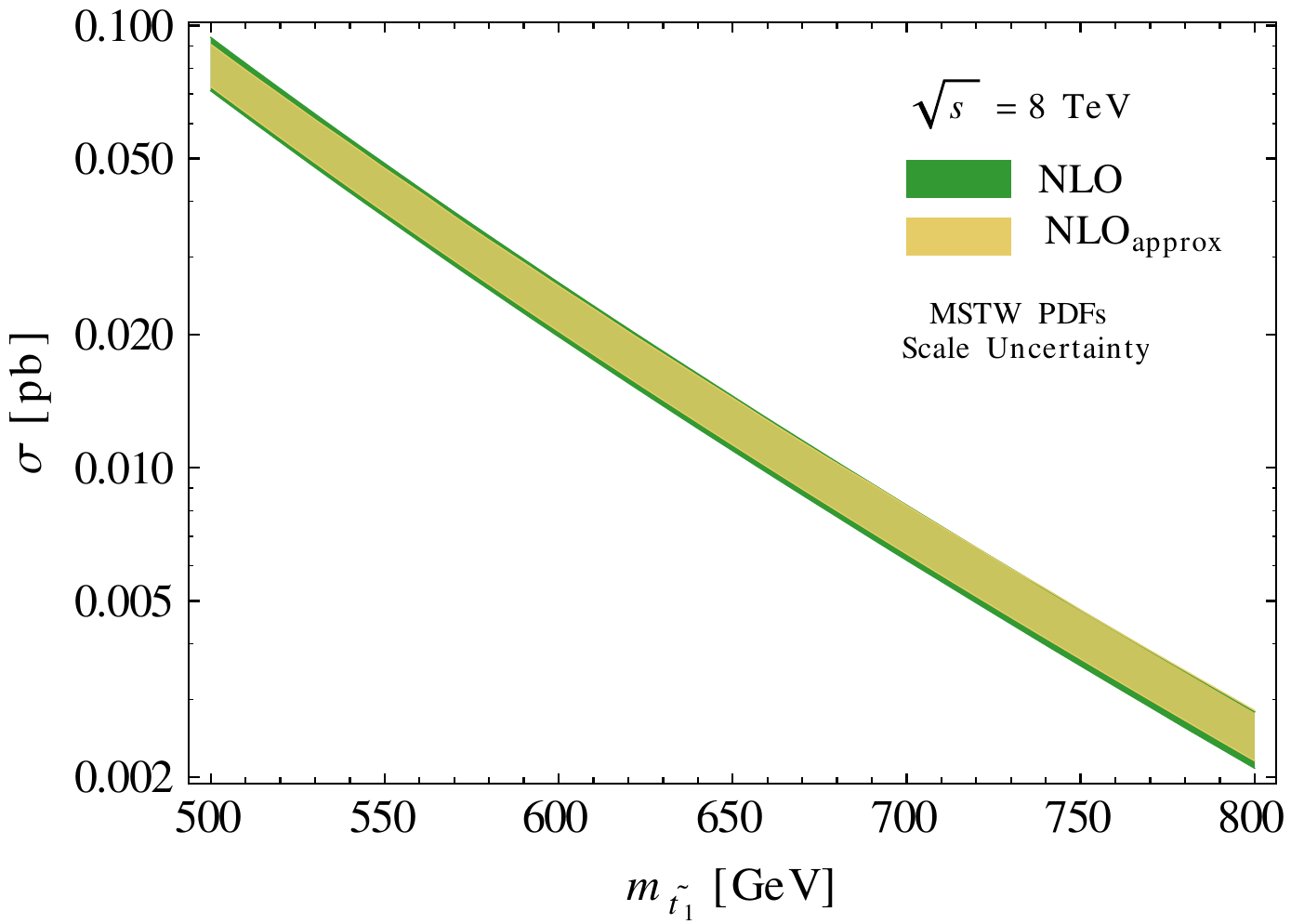} &  \includegraphics[width=0.48\textwidth]{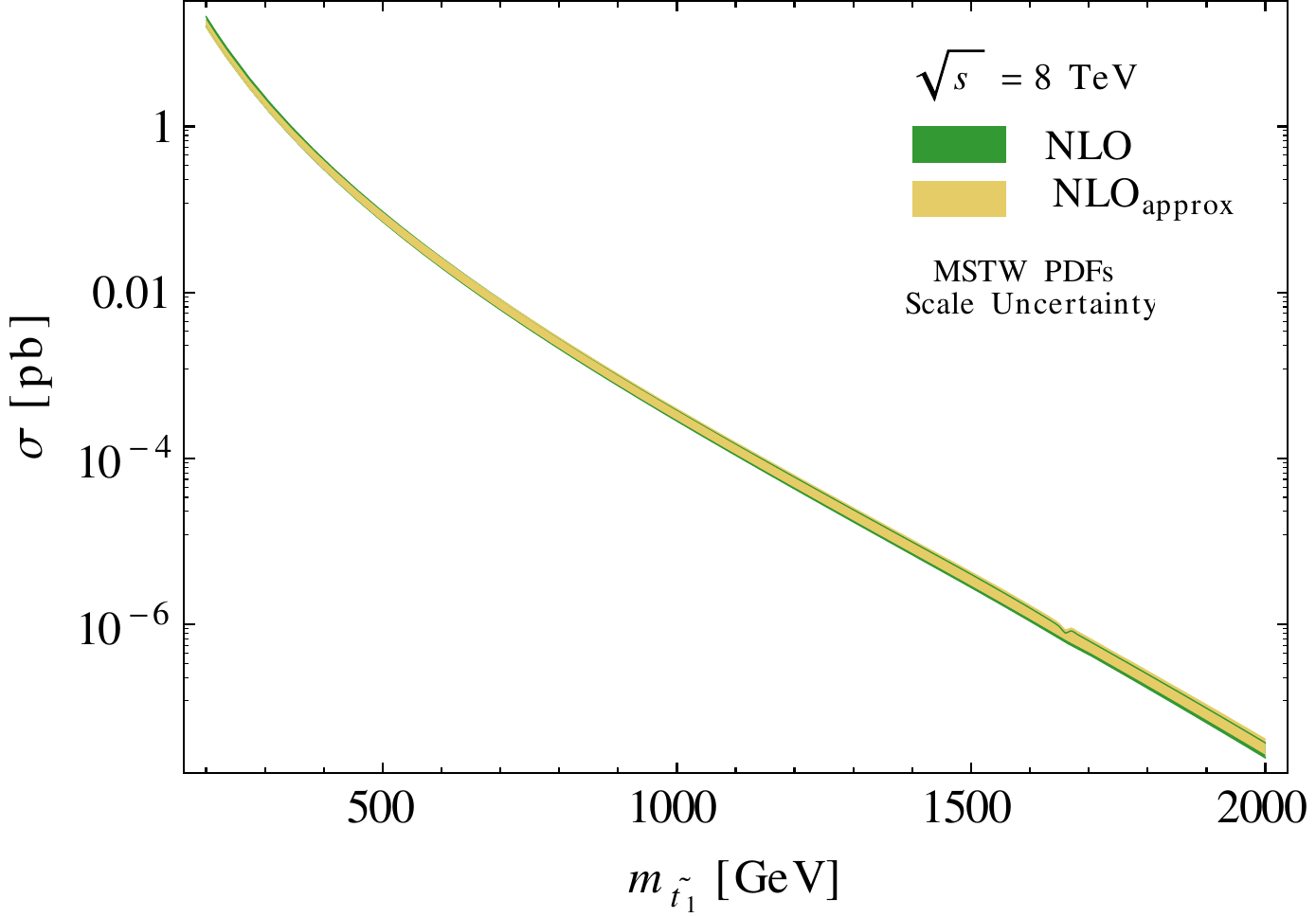}\\ 
\end{tabular} 
\end{center}
\caption{Comparison of the full NLO cross section with the one obtained at approximate NLO. 
All panels refer to the LHC with a hadronic center of mass energy of $8$~TeV. 
In the first and second  row we employ CT10 \cite{Lai:2010vv} and MSTW2008 NLO 
\cite{Martin:2009iq,Martin:2009bu,Martin:2010db} PDFs, respectively. 
The left and right columns show different ranges in the stop mass.
\label{NLOcompX}}
\end{figure}

As a first step, we compare the full NLO cross section with the approximate NLO cross section given by the leading singular terms. To be specific, in the approximate NLO formulas we include the coefficients $D_{1,ij}^{(1)}$, $D_{0,ij}^{(1)}$, and $Q_{0,ij}^{(1)}$ in Eqs.~(\ref{Cpim}) and (\ref{C1pi}), as well as those terms in $R_{ij}^{(1)}$ which naturally arise in the SCET approach.
The purpose of this comparison is to establish to what extent the leading terms in the threshold approximation reproduce the full cross section, or, in other words, if the dynamical threshold enhancement of the soft emission region takes place.  This comparison is shown in Figure~\ref{NLOcompX}, for the case of a hadronic center of mass energy of $8$~TeV. The two rows in the  figure refer to two different choices of the PDF set. NLO PDFs are employed in all of the four panels. One observes that the average of the approximate PIM and 1PI NLO formulas reproduces very well the band obtained by varying the factorization scale (which is set equal to the renormalization scale) in the full NLO result. 
Furthermore, the comparison in Figure~\ref{NLOcompX} supports the fact that the integrals of the differential distributions in PIM and 1PI kinematics over the whole available phase space  reproduce to a good accuracy the total cross section, at least for the range of stop masses of interest in this work.
It is thus reasonable to expect that also the approximate NNLO formulas reproduce to a good extent the unknown full NNLO corrections.


\begin{figure}[t]
\begin{center}

\begin{tabular}{cc}
\includegraphics[width=0.48\textwidth]{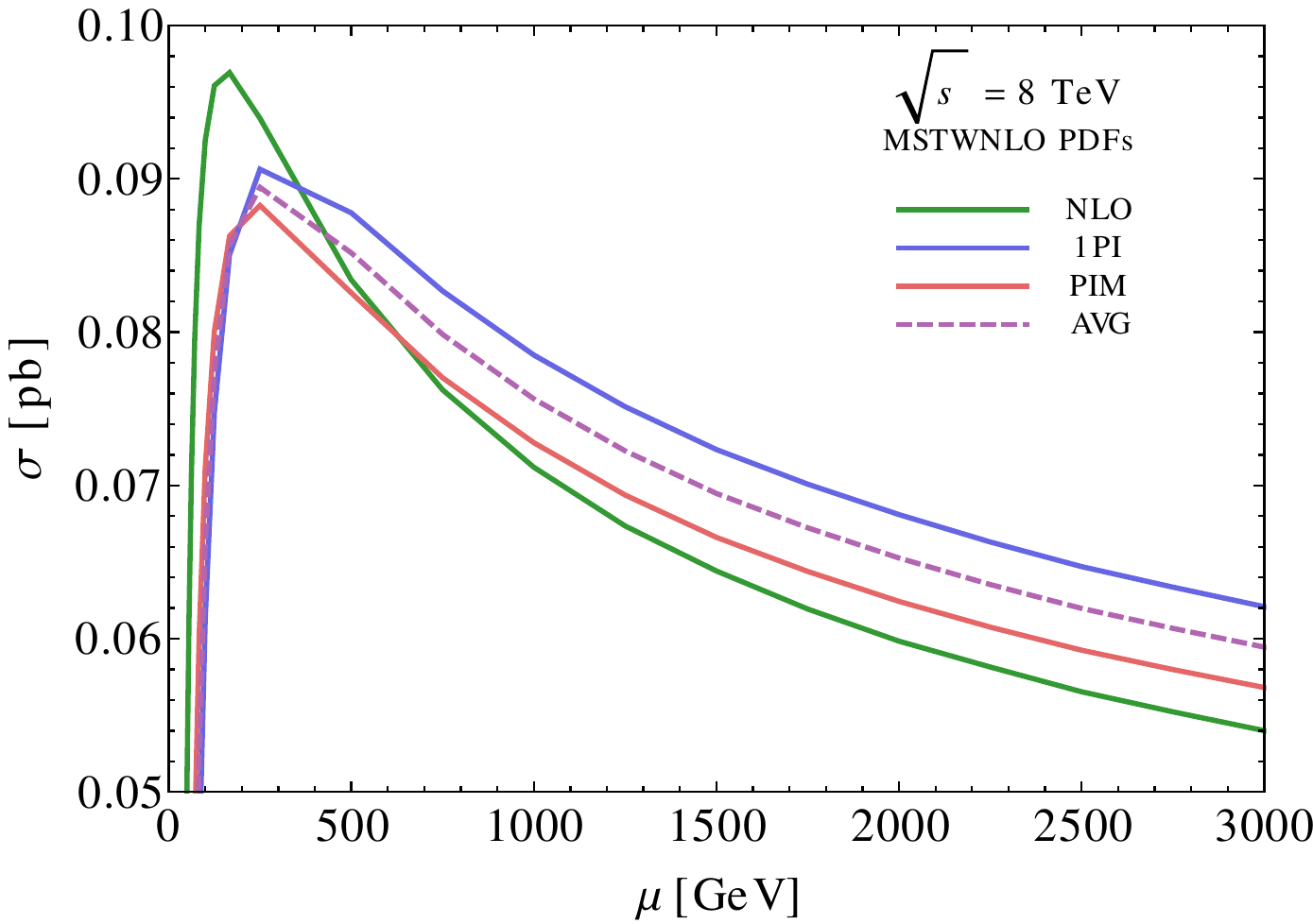}  &  \includegraphics[width=0.463\textwidth]{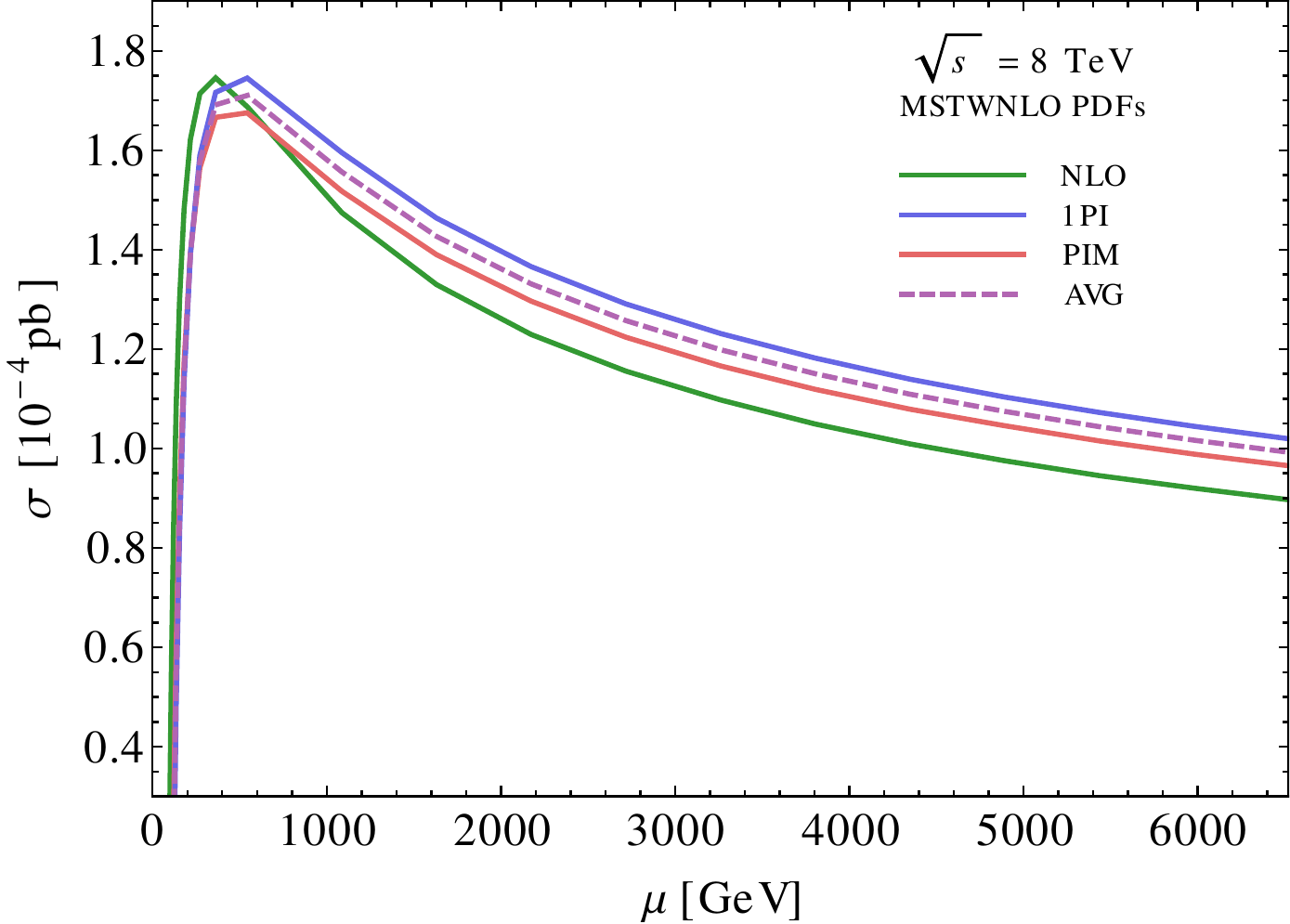} \\ 
\includegraphics[width=0.48\textwidth]{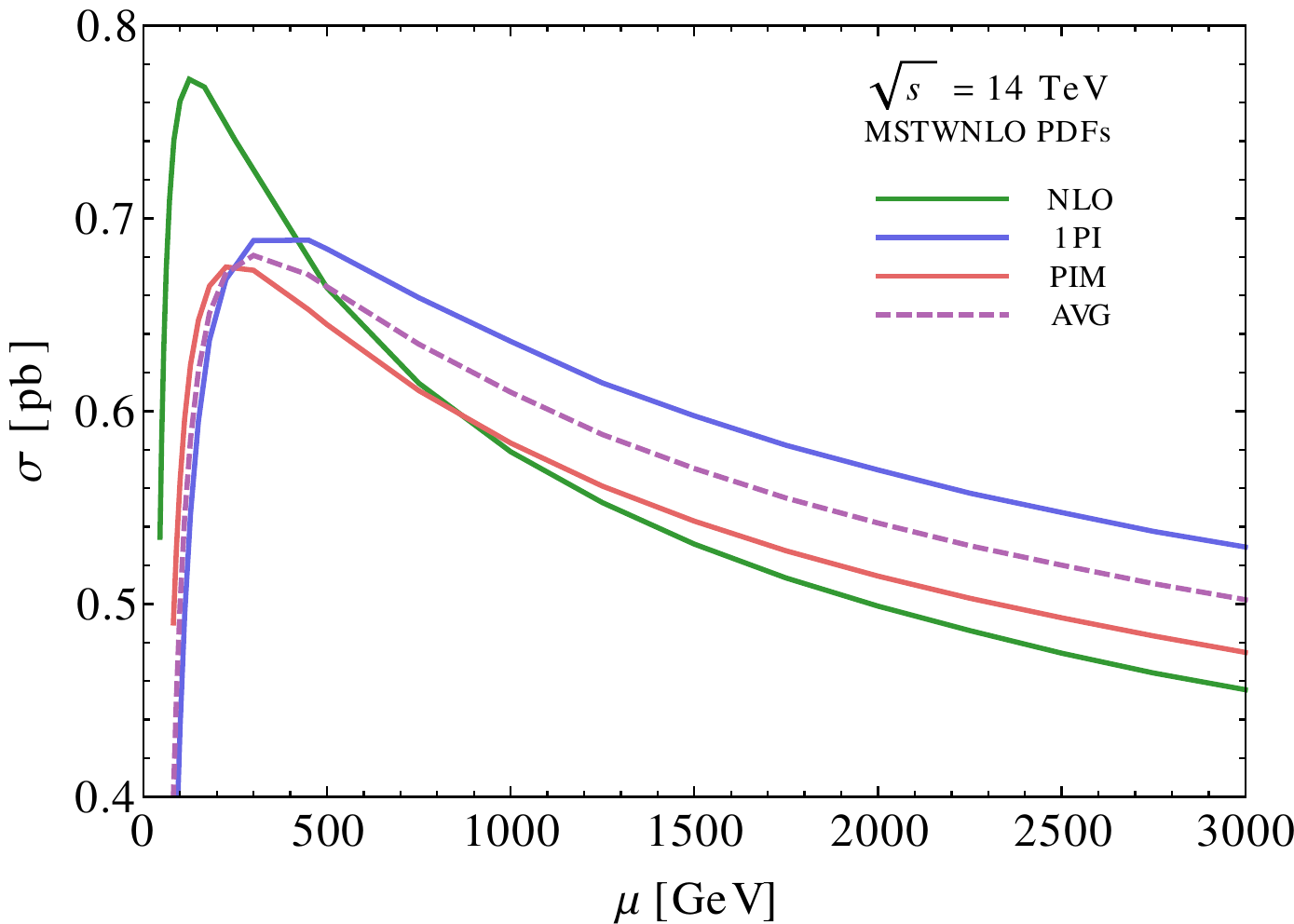}  &  \includegraphics[width=0.465\textwidth]{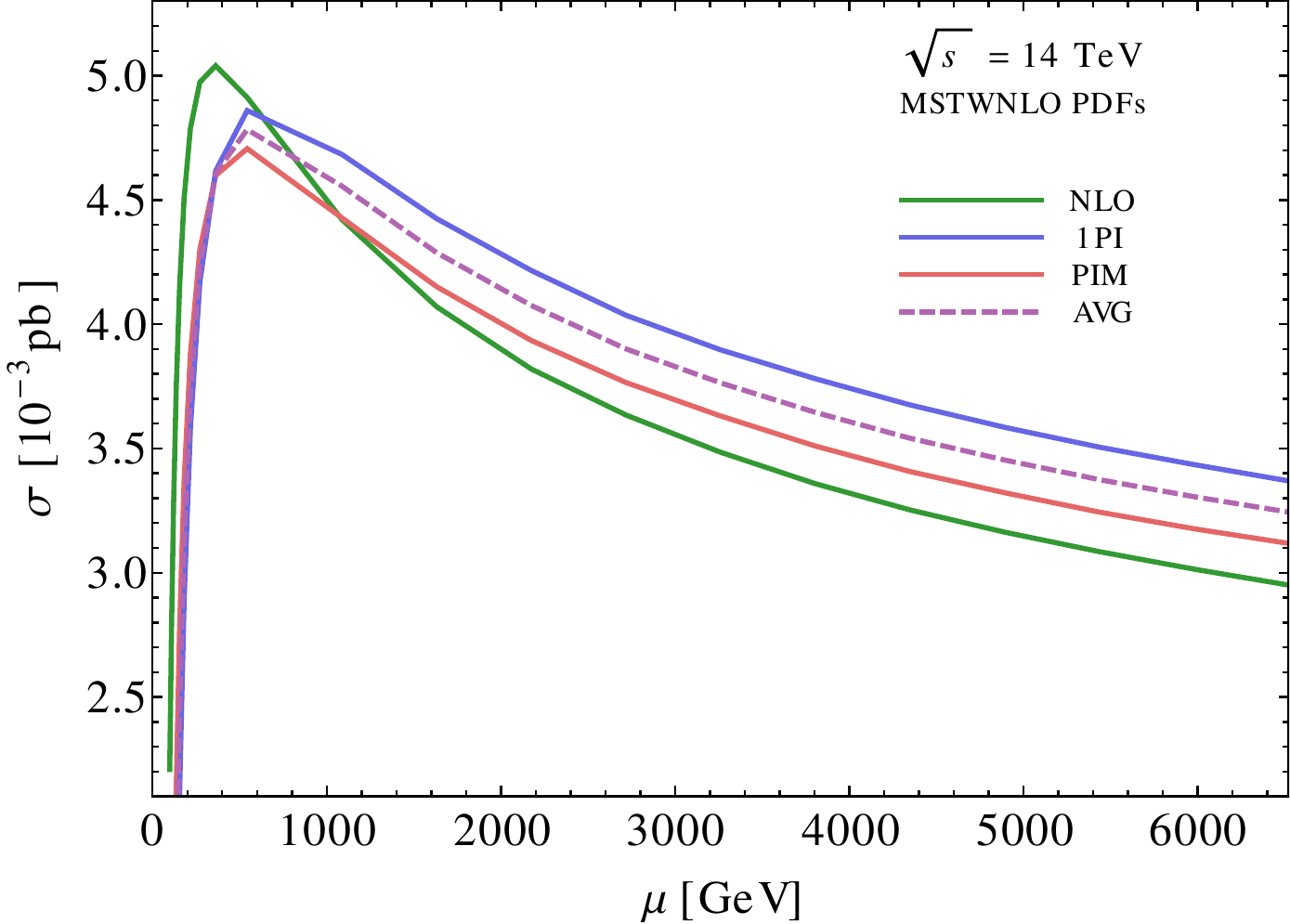}\\  
\end{tabular} 
\end{center}
\caption{Scale dependence of the PIM, 1PI, and PIM-1PI average approximate NLO  cross sections compared to the full NLO cross section. The left panels refer to the production of a top squark of mass $m_{\tilde{t}_1} = 500$~GeV, the right panels refer to the case $m_{\tilde{t}_1} = 1087.17$~GeV. The two figures in the first row are obtained for LHC at $\sqrt{S} = 8$~TeV,
the ones in the second row refer to $\sqrt{S} = 14$~TeV.
All of the SUSY parameters other than $m_{\tilde{t}_1}$ are fixed at the values of the benchmark point {\tt 40.2.5} \cite{AbdusSalam:2011fc}.
\label{scalepim1pi}}
\end{figure}

\begin{figure}[t]
\begin{center}

\begin{tabular}{cc}
\includegraphics[width=0.48\textwidth]{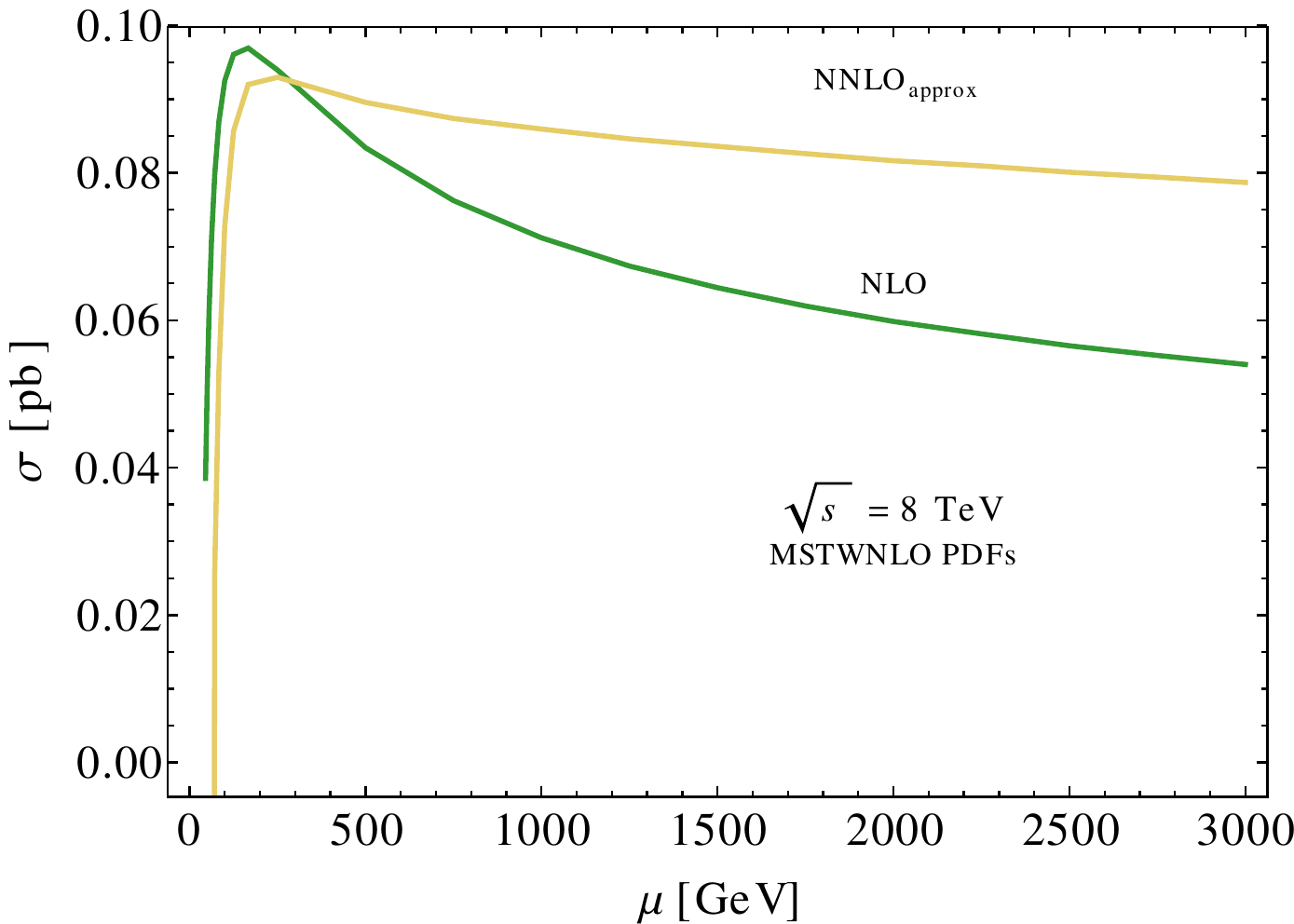}  &  \includegraphics[width=0.465\textwidth]{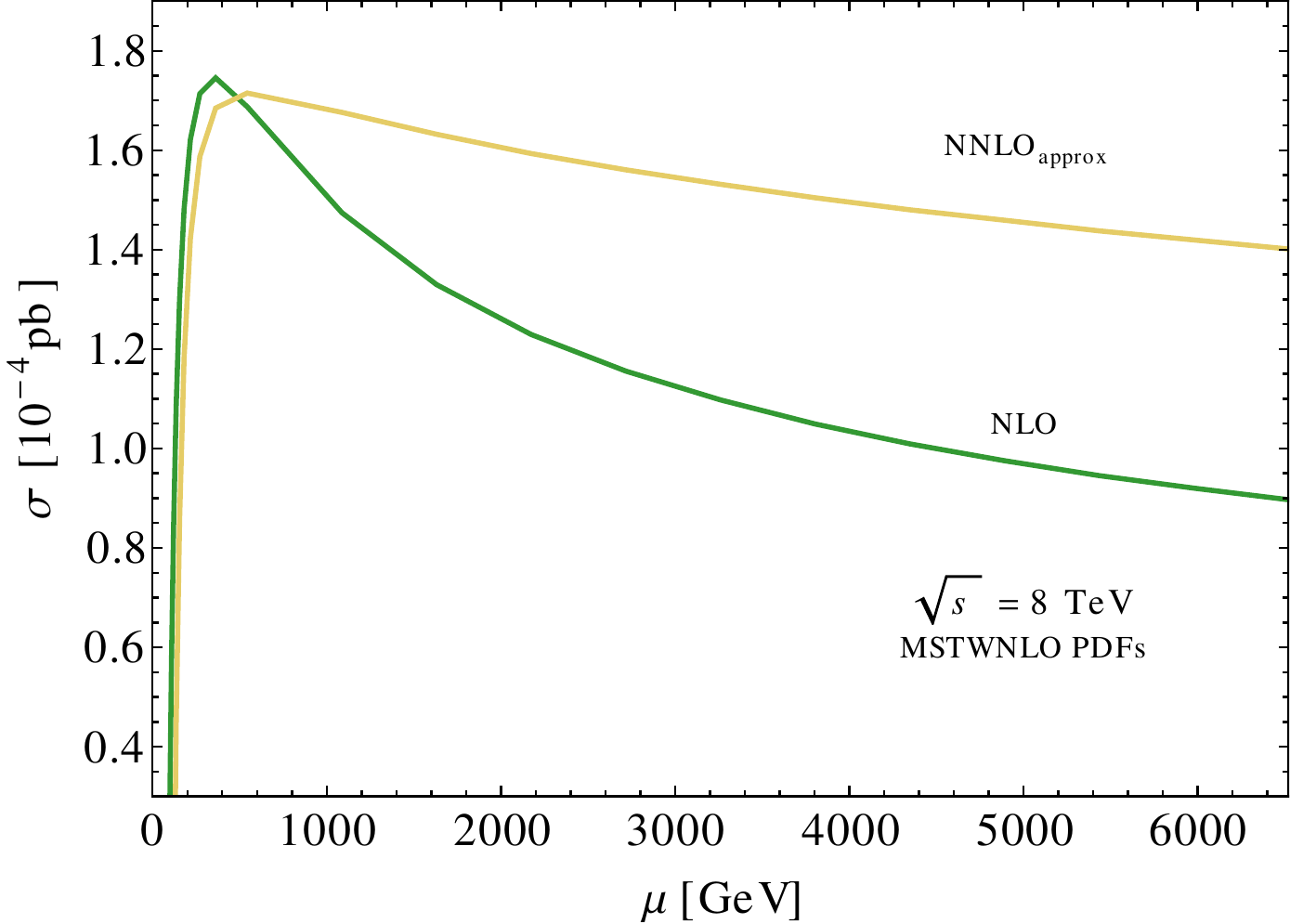} \\ 
\includegraphics[width=0.48\textwidth]{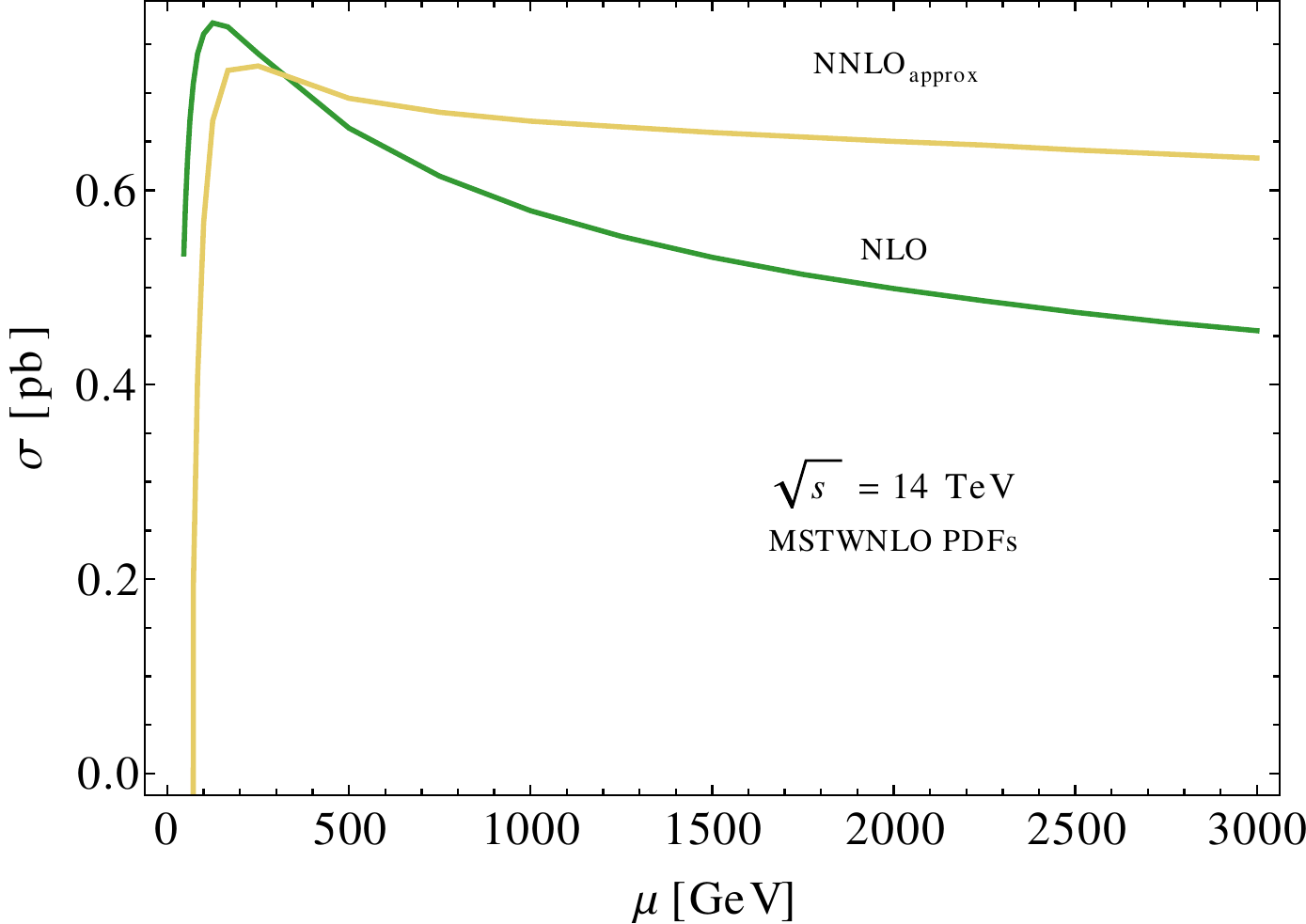}  &  \includegraphics[width=0.465\textwidth]{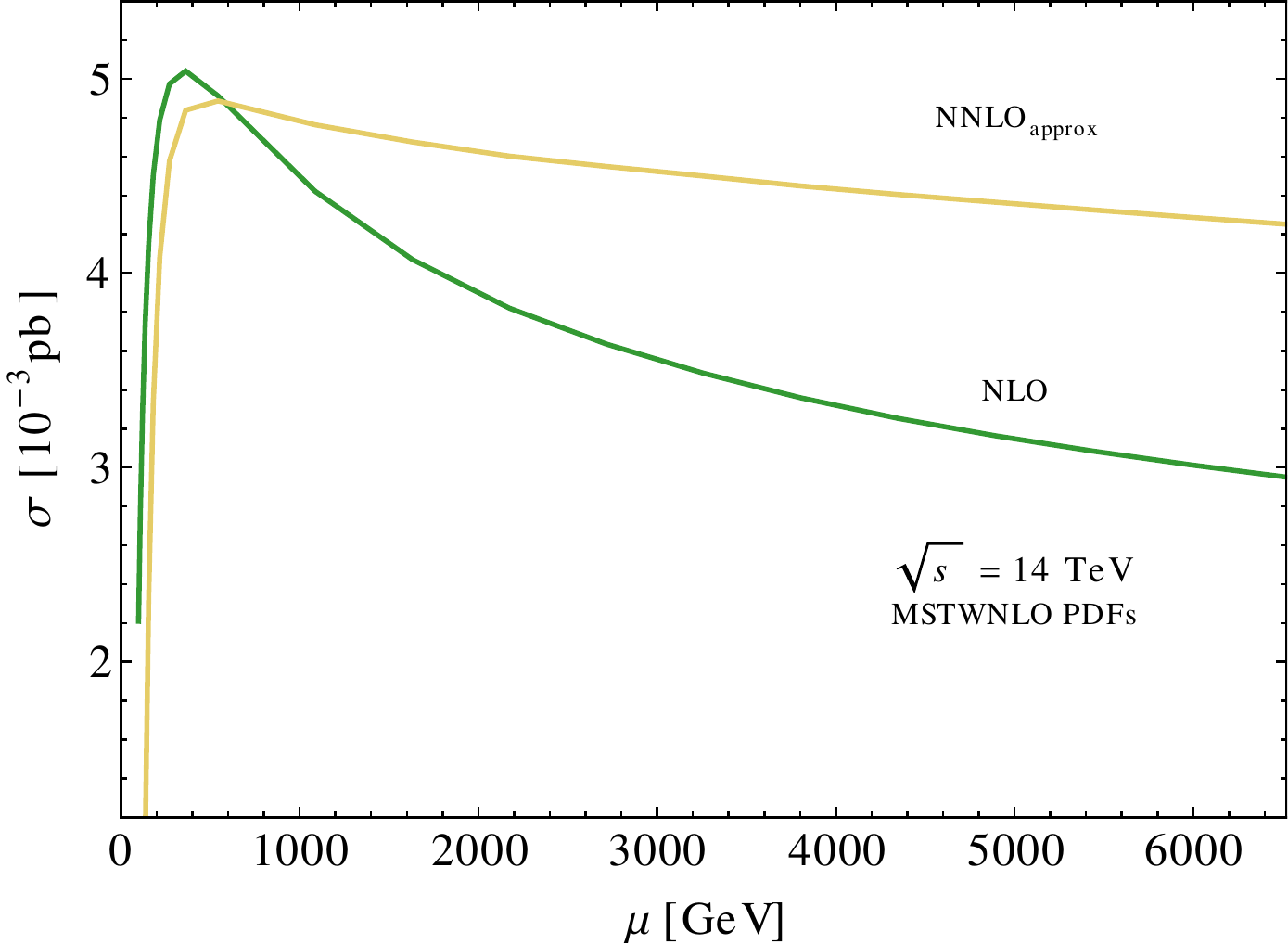}\\  
\end{tabular} 
\end{center}
\caption{Scale dependence of the NLO and approximate NNLO cross sections. The left panels refer to the production of a top squark of mass $m_{\tilde{t}_1} = 500$~GeV, the right panels refer to the case $m_{\tilde{t}_1} = 1087.17$~GeV. The two figures in the first row are obtained for the LHC at $\sqrt{S} = 8$~TeV,
the ones in the second row refer to $\sqrt{S} = 14$~TeV.
All of the SUSY parameters other than $m_{\tilde{t}_1}$ are set to the values of the benchmark point {\tt 40.2.5} \cite{AbdusSalam:2011fc}.
\label{NNLOscale}}
\end{figure}

Another comparison between the scale dependence of the  approximate NLO  cross section and the full NLO cross section is shown in Figure~\ref{scalepim1pi}, in which we plot separate curves for the cross section obtained in $\PIMs$ and $\IPIs$ kinematics, and for the average of the two. Both for $m_{\tilde{t}_1} = 500$~GeV and for $m_{\tilde{t}_1} = 1087.17$~GeV, the scale dependence
of the approximate cross sections in the range $m_{\tilde{t}_1}/2 \le \mu \le 2 m_{\tilde{t}_1}$ reproduces to a good extent the scale variation of the full NLO cross section in the same range.

We now turn to the discussion of  the approximate NNLO results, recalling first that in the approximate NNLO formulas we do not include the terms proportional to the Dirac delta functions $\delta(1-z)$ ($\PIMs$) or $\delta(s_4)$ ($\IPIs$) arising from the NNLO hard functions, since the scale-independent parts of the NNLO hard functions are still unknown. 
As expected, the approximate NNLO predictions for the pair-production cross section show a smaller scale dependence than the NLO results for the same quantity. This is illustrated in Figure~\ref{NNLOscale}, where two different LHC center of mass energies ($8$~TeV and $14$~TeV) and two different stop-quark masses 
($m_{\tilde{t}_1} =  500$~GeV and $m_{\tilde{t}_1} =  1087.17$~GeV) are considered.
The approximate NNLO line is obtained by plotting the scale dependence of the average between the PIM and 1PI cross sections.
In order to show the effect of the approximate NNLO corrections on the scale dependence, both the NLO and approximate NNLO curves are  plotted using MSTW2008 NLO PDF sets. 
While in Figure~\ref{NNLOscale} we plot exclusively the cross section scale dependence, 
we remind the reader that the perturbative uncertainty employed in all tables as well as in all other figures is obtained by evaluating the 
second and third line of Eqs.~(\ref{scalencertainty}). Consequently, it reflects both the scale uncertainty and a  kinematic scheme uncertainty, which is associated to the different sets of non-singular terms neglected in $\IPIs$ and 
$\PIMs$ kinematics.


\begin{table}[t]
\centering
\begin{tabular}{|c||c|c|}
\hline
LHC $7$~TeV &\multicolumn{2}{|c|}{MSTW2008} \\
\hline
$m_{\st}$~[GeV] & 500 & 1087.17 \\
\hline
\hline
$(\sigma \pm \Delta \sigma_\mu \pm \Delta_{\rm{pdf} })_{\rm{{\tiny LO}}}$ [pb] &$34.4^{+15.8+ 3.8 }_{-10.0-3.6} \times 10^{-3}$  & $38.8^{+19.7+10.4 }_{-12.1 -8.2 } \times 10^{-6}$ \\
\hline
$(\sigma \pm \Delta \sigma_\mu \pm \Delta_{\rm{pdf} + \alpha_s})_{\rm{{\tiny NLO}}}$ [pb] & $46.2^{+6.1+6.6}_{-7.0-5.3} \times 10^{-3}$  &$49.6^{+8.0+15.2}_{-8.7-10.5} \times 10^{-6} $ \\
\hline
$(\sigma \pm \Delta \sigma_\mu\pm \Delta_{\rm{pdf} + \alpha_s})_{\rm{{\tiny approx.\ NNLO}}}$ [pb] &$46.2^{+1.7+8.0}_{-2.9-5.9} \times 10^{-3}$  &$52.8^{+1.4+25.8}_{-3.6-12.1} \times 10^{-6}$ \\
\hline
$K_{\rm{NLO}}$ & $1.34$  &$1.28$  \\
\hline
$K_{\rm{approx.\ NNLO}}$ & $1.34$  & $1.36$ \\
\hline
\end{tabular}
\caption{Stop-pair production cross section for two different values of $m_{\tilde{t}_1}$ at the LHC with  $\sqrt{S} = 7$~TeV. The numbers are obtained by using MSTW2008 PDFs. Here and in the following tables, all of the SUSY parameters (with the exception of $m_{\tilde{t}_1}$) are fixed at the values prescribed by the benchmark point {\tt 40.2.5} \cite{AbdusSalam:2011fc}.\label{tab:7tevMSTW.40.2.5}}
\end{table}

\begin{table}[t]
\centering
\begin{tabular}{|c||c|c|}
\hline
LHC $7$~TeV  &\multicolumn{2}{|c|}{CT10}  \\
\hline
$m_{\st}$~[GeV] & 500 & 1087.17 \\
\hline
\hline
$(\sigma \pm \Delta \sigma_\mu \pm \Delta_{\rm{pdf}+ \alpha_s})_{\rm{{\tiny LO}}}$ [pb] &$30.1^{+12.2+7.1}_{-8.1-5.1 }\times 10^{-3}$  & $36.7^{+17.9 + 30.7}_{-11.3-13.4} \times 10^{-6}$\\
\hline
$(\sigma \pm \Delta \sigma_\mu \pm \Delta_{\rm{pdf} + \alpha_s})_{\rm{{\tiny NLO}}}$ [pb] & $ 45.3^{+5.8+11.0}_{-6.6-8.1}\times 10^{-3}$  &$58.4^{+9.3+49.9}_{-10.2-22.4} \times 10^{-6} $ \\
\hline
$(\sigma \pm \Delta \sigma_\mu\pm \Delta_{\rm{pdf} + \alpha_s})_{\rm{{\tiny approx.\ NNLO}}}$ [pb] &$46.7^{+1.7+11.6}_{-2.9-8.3} \times 10^{-3}$  &$51.7^{+1.1+34.1}_{-3.6-18.6} \times 10^{-6}$  \\
\hline
$K_{\rm{NLO}}$ &$1.50$  & $1.59$\\
\hline
$K_{\rm{approx.\ NNLO}}$ & $1.55$ & $1.41$ \\
\hline
\end{tabular}
\caption{Stop-pair production cross section for two different values of $m_{\tilde{t}_1}$ at the  LHC with $\sqrt{S} = 7$~TeV. The numbers are obtained by using CT10 PDFs.
\label{tab:7tevCT10.40.2.5}}
\end{table}

\begin{table}[th]
\centering
\begin{tabular}{|c||c|c|}
\hline
LHC $8$~TeV &\multicolumn{2}{|c|}{MSTW2008}  \\
\hline
$m_{\st}$~[GeV] & 500 & 1087.17 \\
\hline
\hline
$(\sigma \pm \Delta \sigma_\mu \pm \Delta_{\rm{pdf} })_{\rm{{\tiny LO}}}$ [pb] & $61.7^{+27.3 + 6.1 }_{-17.5 -6.0}  \times 10^{-3}$  & $11.5^{+5.6 + 2.5 }_{-3.5 -2.0}\times 10^{-5}$\\
\hline
$(\sigma \pm \Delta \sigma_\mu \pm \Delta_{\rm{pdf} + \alpha_s})_{\rm{{\tiny NLO}}}$ [pb] &$83.4^{+10.5 + 10.6}_{-12.2 -8.8} \times 10^{-3}$  & $ 14.7^{+2.1+ 3.7}_{-2.5 -2.8}\times 10^{-5}$ \\
\hline
$(\sigma \pm \Delta \sigma_\mu\pm \Delta_{\rm{pdf} + \alpha_s})_{\rm{{\tiny approx.\ NNLO}}}$ [pb] & $ 83.2^{+3.3 + 12.6}_{-4.9-9.9} \times 10^{-3}$ & $ 15.3^{+0.3+5.8}_{-1.0-3.0} \times 10^{-5}$ \\
\hline
$K_{\rm{NLO}}$ & $1.35$ & $1.29$  \\
\hline
$K_{\rm{approx.\ NNLO}}$ & $1.35$ & $1.34$ \\
\hline
\end{tabular}
\caption{Stop-pair production cross section for two different values of $m_{\tilde{t}_1}$ at the LHC with $\sqrt{S} = 8$~TeV. The numbers are obtained by using MSTW2008 PDFs.
\label{tab:8tevMSTW.40.2.5}}
\end{table}

\begin{table}[th]
\centering
\begin{tabular}{|c||c|c|}
\hline
LHC $8$~TeV &\multicolumn{2}{|c|}{CT10}  \\
\hline
$m_{\st}$~[GeV] & 500 & 1087.17 \\
\hline
\hline
$(\sigma \pm \Delta \sigma_\mu \pm \Delta_{\rm{pdf} + \alpha_s})_{\rm{{\tiny LO}}}$ [pb] & $54.0^{+21.2 + 11.0}_{-14.2 -8.3} \times 10^{-3}$ & $10.6^{+4.8 +6.6}_{-3.1-3.2}\times 10^{-5}$ \\
\hline
$(\sigma \pm \Delta \sigma_\mu \pm \Delta_{\rm{pdf} + \alpha_s})_{\rm{{\tiny NLO}}}$ [pb] & $80.9^{+9.8 + 16.6}_{-11.4 -13.1} \times 10^{-3}$  & $ 16.5^{+2.3+ 10.4}_{-2.7-5.3} \times 10^{-5}$ \\
\hline
$(\sigma \pm \Delta \sigma_\mu\pm \Delta_{\rm{pdf} + \alpha_s})_{\rm{{\tiny approx.\ NNLO}}}$ [pb] & $ 83.6^{+3.6+ 19.0}_{-4.8-12.3} \times 10^{-3}$ &  $ 15.2^{+0.3+ 8.1}_{-1.0-4.7} \times 10^{-5}$ \\
\hline
$K_{\rm{NLO}}$ &$1.50$  & $1.56$  \\
\hline
$K_{\rm{approx.\ NNLO}}$ &$1.55$  & $1.44$  \\
\hline
\end{tabular}
\caption{Stop-pair production cross section for two different values of $m_{\tilde{t}_1}$ at the LHC with $\sqrt{S} = 8$~TeV. The numbers are obtained by using CT10 PDFs.
\label{tab:8tevCT10.40.2.5}}
\end{table}

\begin{table}[th]
\centering
\begin{tabular}{|c||c|c|}
\hline
LHC $14$~TeV &\multicolumn{2}{|c|}{MSTW2008} \\
\hline
$m_{\st}$~[GeV] & 500 & 1087.17 \\
\hline
\hline
$(\sigma \pm \Delta \sigma_\mu \pm \Delta_{\rm{pdf} })_{\rm{{\tiny LO}}}$ [pb] & $48.3^{+18.4+3.3}_{-12.4-3.4} \times 10^{-2}$ & $ 33.5^{+13.8+3.7}_{-9.1-3.6} \times 10^{-4}$\\
\hline
$(\sigma \pm \Delta \sigma_\mu \pm \Delta_{\rm{pdf} + \alpha_s})_{\rm{{\tiny NLO}}}$ [pb] & $66.4^{+7.7+6.2}_{-8.5-5.2} \times 10^{-2}$ & $ 44.2^{+4.9+6.4}_{-6.0-5.1} \times 10^{-4}$ \\
\hline
$(\sigma \pm \Delta \sigma_\mu\pm \Delta_{\rm{pdf} + \alpha_s})_{\rm{{\tiny approx.\ NNLO}}}$ [pb] & $65.7^{+3.3+6.5}_{-3.4-6.2}\times 10^{-2}$ & $ 44.3^{+1.3+7.8}_{-2.2-5.4} \times 10^{-4}$  \\
\hline
$K_{\rm{NLO}}$ & $1.38$  & $1.32$  \\
\hline
$K_{\rm{approx.\ NNLO}}$ & $1.36$ &  $1.32$\\
\hline
\end{tabular}
\caption{Stop-pair production cross section for two different values of $m_{\tilde{t}_1}$ at the LHC with $\sqrt{S} = 14$~TeV. The numbers are obtained by using MSTW2008 PDFs.
\label{tab:14tevMSTW.40.2.5}}
\end{table}

\begin{table}[th]
\centering
\begin{tabular}{|c||c|c||}
\hline
LHC $14$~TeV &\multicolumn{2}{|c|}{CT10}  \\
\hline
$m_{\st}$~[GeV] & 500 & 1087.17 \\
\hline
\hline
$(\sigma \pm \Delta \sigma_\mu \pm \Delta_{\rm{pdf} + \alpha_s})_{\rm{{\tiny LO}}}$ [pb] & $42.6^{+14.4+5.0}_{-10.1-4.3} \times 10^{-2}$ & $ 30.1^{+11.3+ 7.8}_{-7.7 -5.2} \times 10^{-4}$ \\
\hline
$(\sigma \pm \Delta \sigma_\mu \pm \Delta_{\rm{pdf} + \alpha_s})_{\rm{{\tiny NLO}}}$ [pb] & $63.2^{+7.0+7.6}_{-7.8-6.6} \times 10^{-2}$ & $ 44.1^{+4.8+11.7}_{-5.8-8.1} \times 10^{-4}$ \\
\hline
$(\sigma \pm \Delta \sigma_\mu\pm \Delta_{\rm{pdf} + \alpha_s})_{\rm{{\tiny approx.\ NNLO}}}$ [pb] & $65.9^{+3.4+8.2}_{-3.4-6.6} \times 10^{-2}$ & $ 44.6^{+1.3 + 12.1}_{-2.1-7.8} \times 10^{-4}$  \\
\hline
$K_{\rm{NLO}}$ & $1.48$ & $1.47$  \\
\hline
$K_{\rm{approx.\ NNLO}}$ & $1.55$ & $1.48$  \\
\hline
\end{tabular}
\caption{Stop-pair production cross section for two different values of $m_{\tilde{t}_1}$ at the LHC with  $\sqrt{S} = 14$~TeV. The numbers are obtained by using CT10 PDFs.
\label{tab:14tevCT10.40.2.5}}
\end{table}

A more precise assessment of the impact of the approximate NNLO corrections on the central value of the cross section and on the associated perturbative uncertainty can be obtained by comparing predictions for fixed values of the stop mass. This analysis is presented in Tables~\ref{tab:7tevMSTW.40.2.5} and \ref{tab:7tevCT10.40.2.5}, \ref{tab:8tevMSTW.40.2.5}  and  \ref{tab:8tevCT10.40.2.5}, and \ref{tab:14tevMSTW.40.2.5} and  \ref{tab:14tevCT10.40.2.5}, which refer to the LHC with center of mass energies $\sqrt{S}=7$~TeV, 8~TeV, and 14~TeV, respectively. In all tables we consider two different values of the lightest top-squark mass: \emph{i)} the value associated to the benchmark point {\tt 40.2.5}, $m_{\tilde{t}_1} = 1087.17$~GeV, and \emph{ii)} a stop mass $m_{\tilde{t}_1} = 500$~GeV, close to the current lower bound for this particle as determined by searches at the LHC. 
In all tables, the first uncertainty is perturbative, determined  as explained in Eq.~(\ref{scalencertainty}), while the second uncertainty is obtained by scanning over the 90\% CL sets of the corresponding PDFs and by taking into account the error on $\alpha_s(m_Z)$.
In particular, the PDF uncertainty is determined by seeing how the average of the PIM and 1PI predictions changes when evaluated with different members of the PDF set.
The numbers for the cross sections have been obtained by employing PDFs fitted at the corresponding order: LO predictions are obtained by employing LO PDFs, NLO predictions employ NLO PDFs, and approximate NNLO predictions employ NNLO PDFs.

In all cases listed in the tables, the inclusion of the approximate NNLO corrections reduces the perturbative uncertainty, when expressed as a percentage of the central value,  by more than a factor of 2 with respect to the corresponding NLO prediction. We can summarize the content of the tables as follows: the scale variation in the range $m_{\tilde{t}_1}/2 \le \mu \le 2 m_{\tilde{t}_1}$ can increase the NLO central value up to $+[11,16] \%$ or lower it up to $-[13,18] \%$. At approximate NNLO, the scale variation can increase the central value of the cross section up to $+[2,5] \%$ or decrease it up to $- [5,7] \%$. These considerations are valid both when one employs CT10 PDFs \cite{Lai:2010vv} or MSTW2008 PDFs.


\begin{figure}[t]
\begin{center}
\includegraphics[width=0.48\textwidth]{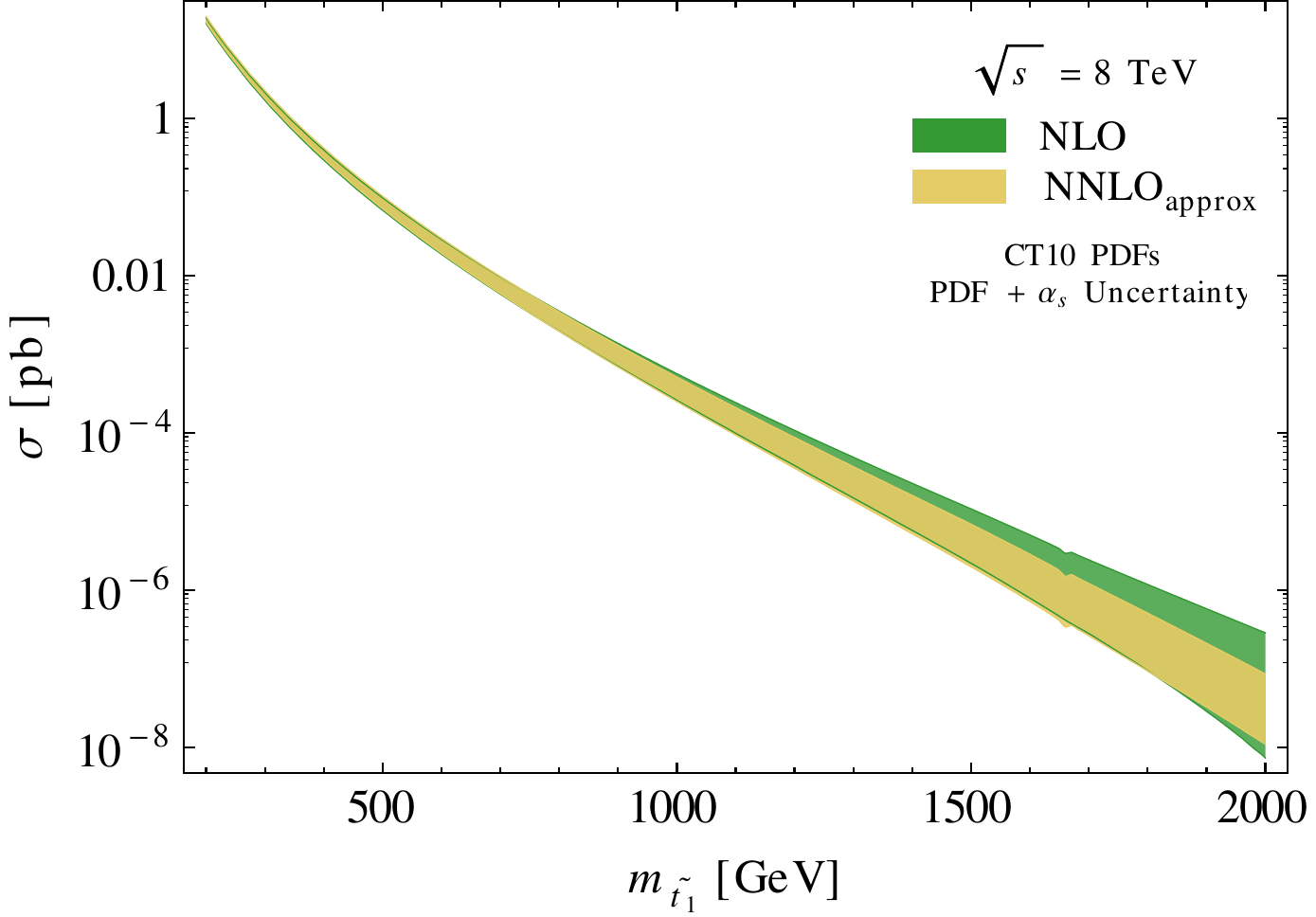}
\quad
\includegraphics[width=0.48\textwidth]{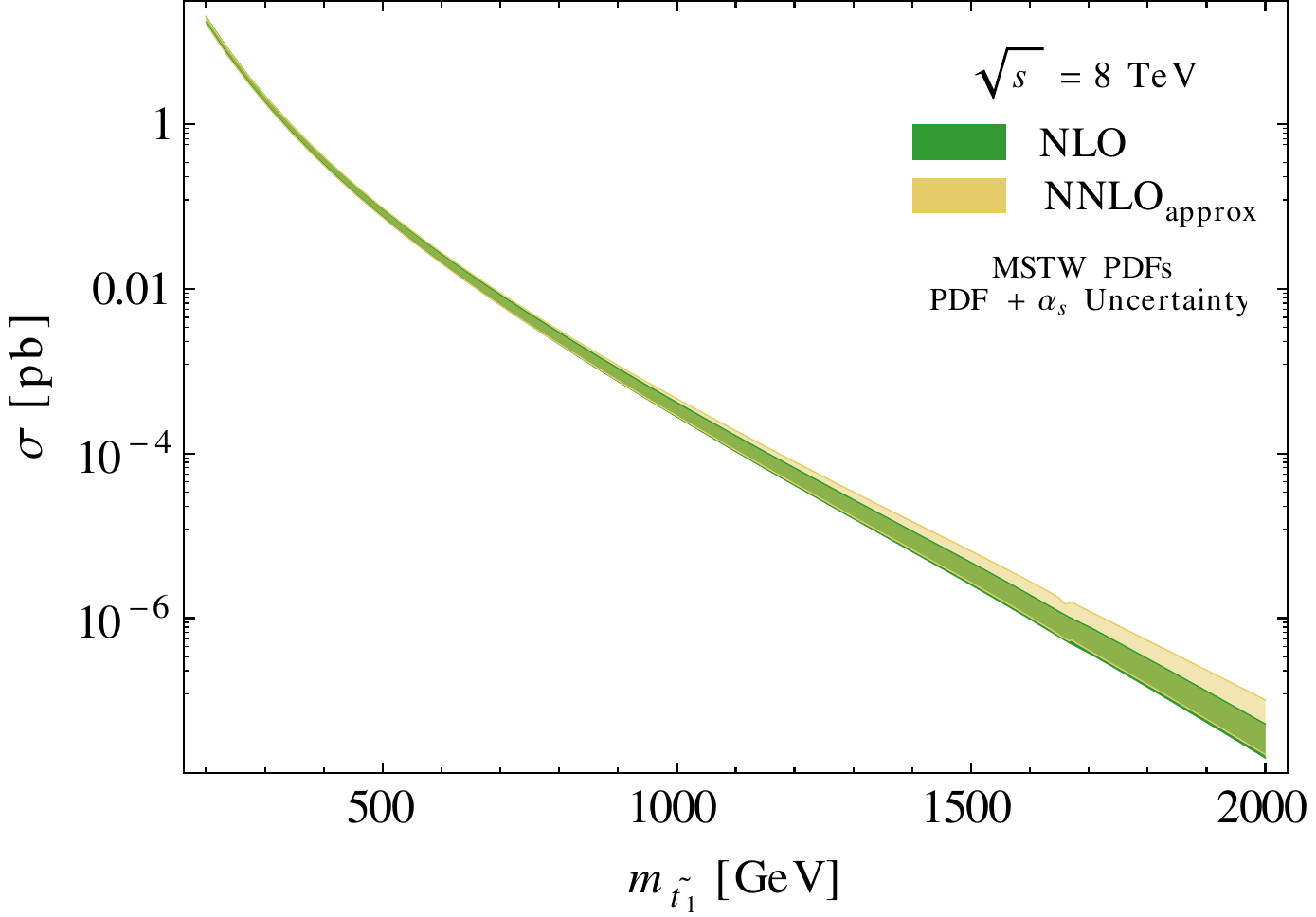}
\end{center}
\caption{PDF and $\alpha_s$ uncertainties on the total stop production cross section at the  LHC with $\sqrt{S} = 8$ TeV. 
\label{PDFanda}}
\end{figure}

In almost all cases illustrated in the tables, the PDF and $\alpha_s$ uncertainty grows marginally in the approximate NNLO predictions with respect to the NLO predictions. Another way to look at the PDF and $\alpha_s$ uncertainty is shown in Figure~\ref{PDFanda}, where this uncertainty band is plotted as function
of the top squark mass in the range $m_{\tilde{t}_1} \in [200,2000]$~GeV at the LHC with center of mass  energy of $8$~TeV. The left panel refers to the case in which CT10 PDFs are employed, while the right panel refers to the case in which the PDFs employed are MSTW2008. One sees that both bands become larger for large stop masses. The approximate NNLO band in the left panel is almost everywhere inside the NLO band, while in the right panel the approximate NNLO band is larger than the NLO band. However, the bands obtained by using CT10 PDFs remain larger than the ones obtained when using MSTW2008 PDFs.


Tables~\ref{tab:7tevMSTW.40.2.5} to~\ref{tab:14tevCT10.40.2.5} also include the values for the $K$ factors at NLO and approximate NNLO, which are both normalized to the LO cross section, i.e.\
\begin{equation}
K_{\rm{NLO}} = \frac{\sigma_{\rm{NLO}}}{\sigma_{\rm{LO}}} \, , \qquad K_{\rm{approx.\ NNLO}} = \frac{\sigma_{\rm{approx. NNLO}}}{\sigma_{\rm{LO}}} \,.
\end{equation}
The NLO $K$ factors tend to be slightly larger when CT10 PDFs rather than MSTW2008 PDFs are employed (roughly 1.5 vs.\ 1.3), but they are not very sensitive to the collider center of mass energy or to the mass of the top squark. The ratio $K_{\rm{approx.\ NNLO}}/K_{\rm{NLO}}$ ranges from $0.88$ to $1.06$, therefore the approximate NNLO corrections have only a moderate impact on the central value of the NLO cross section.\footnote{In this numerical analysis we use NNLO PDFs together with our approximate NNLO results for the hard-scattering kernels. One could also make a different choice and use NLO PDFs with the approximate NNLO formulas, in that case the impact on the central values of our predictions would be bigger.} For top-squark masses smaller than $\sim 1$~TeV,  the central value for the approximate NNLO cross section falls well within the NLO scale uncertainty band.


\begin{figure}[ph]
\begin{center}
\begin{tabular}{cc}
 \includegraphics[width=0.48\textwidth]{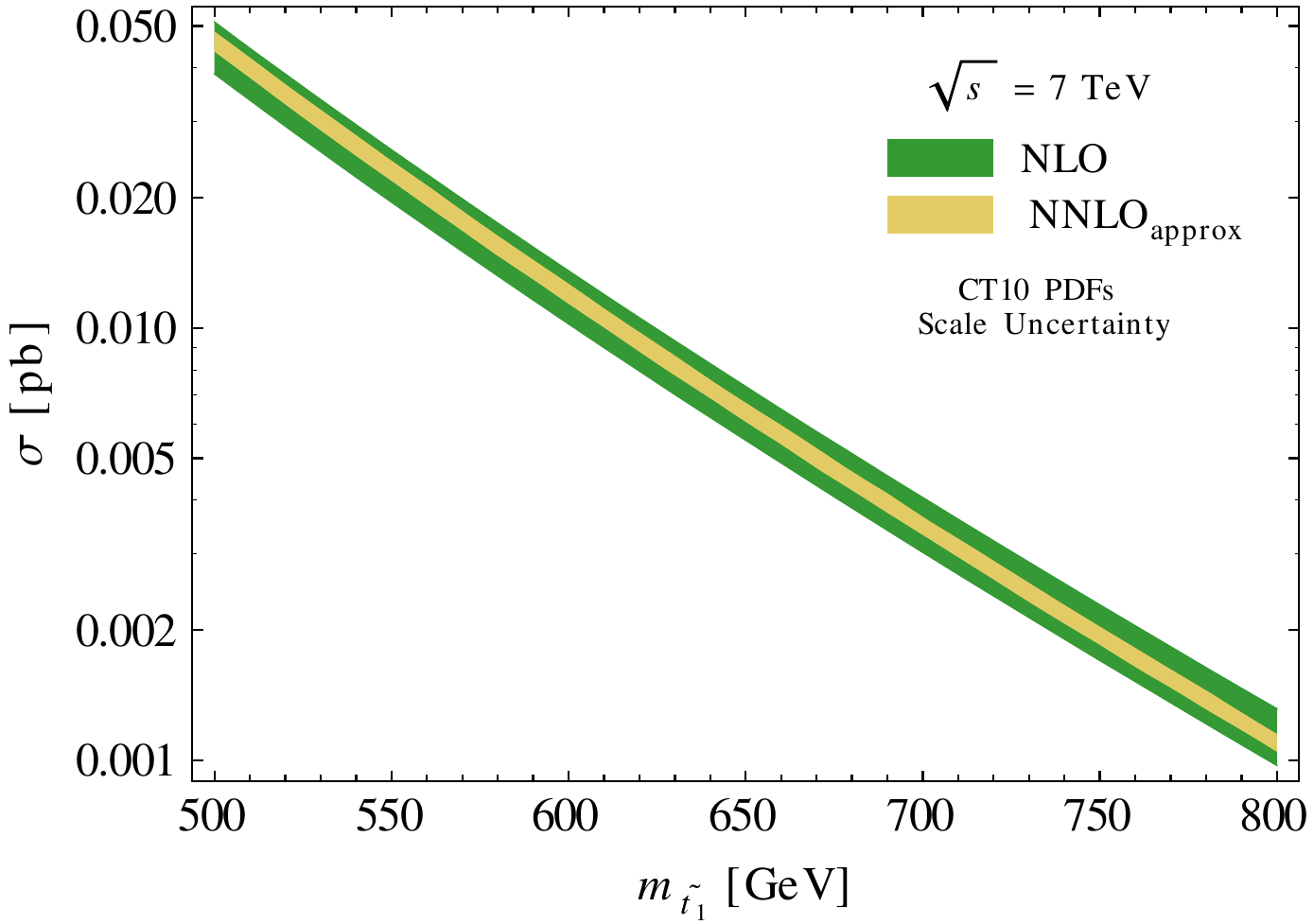} & \includegraphics[width=0.48\textwidth]{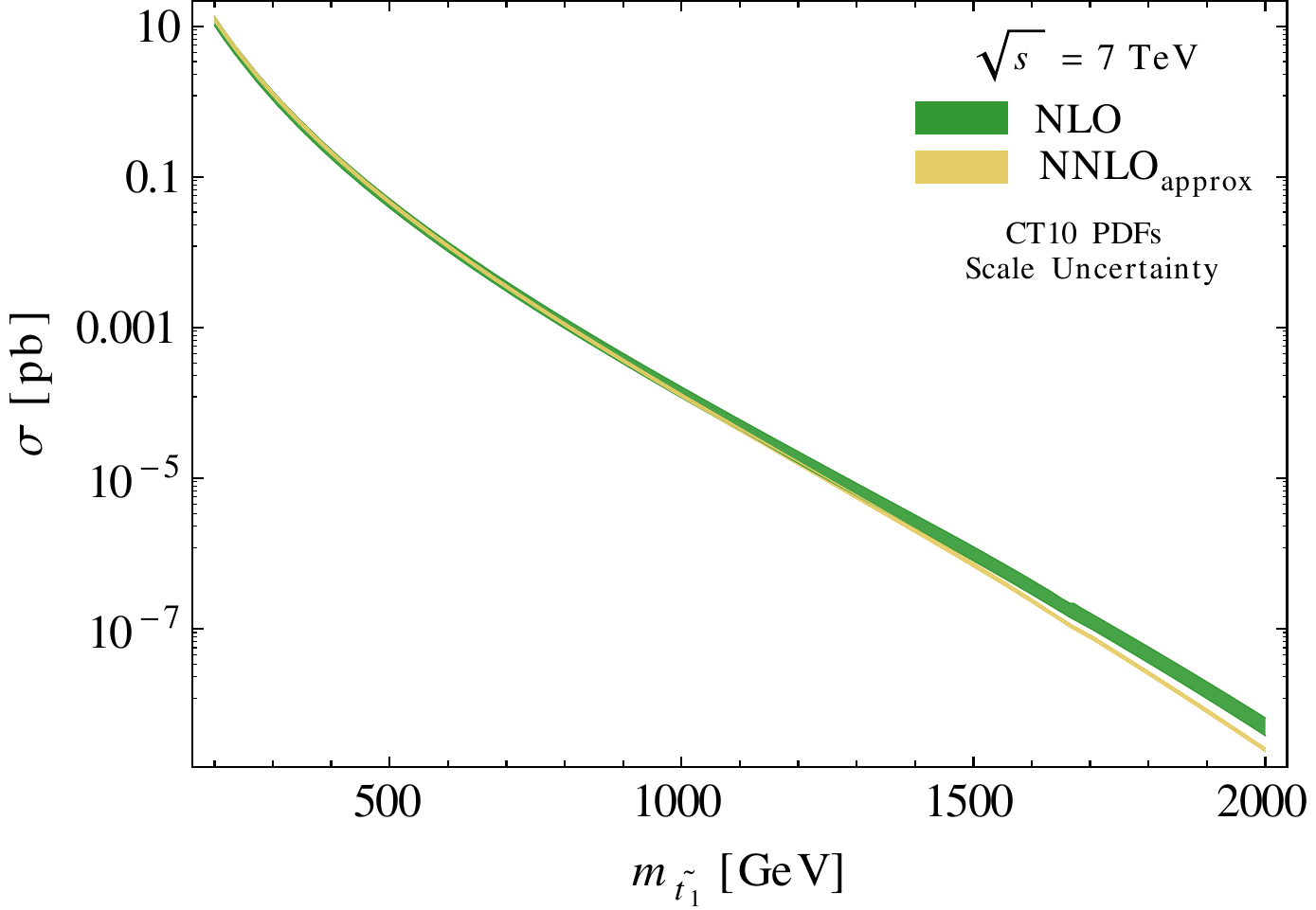} \\ 
 \includegraphics[width=0.48\textwidth]{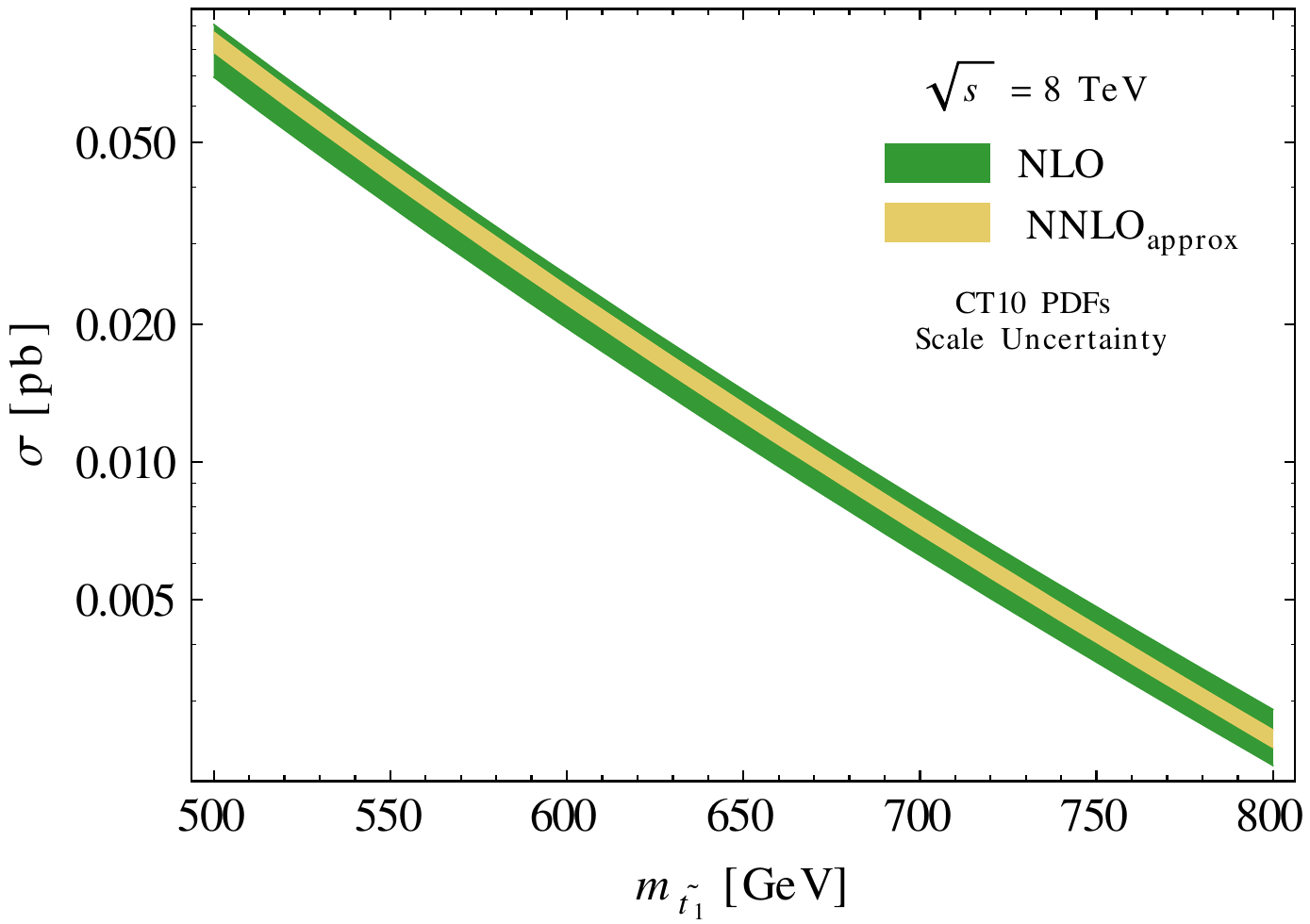} & \includegraphics[width=0.48\textwidth]{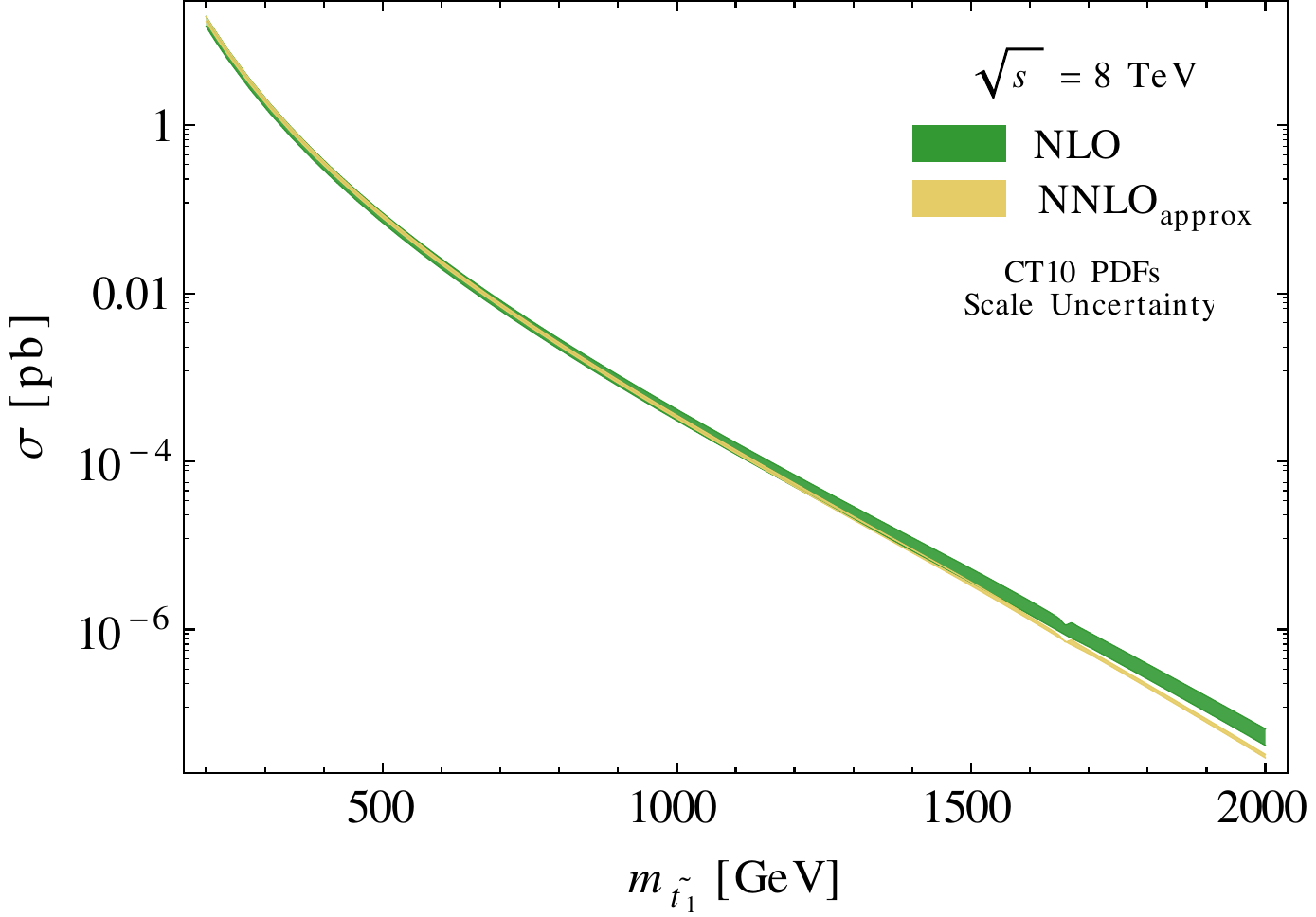} \\ 
 \includegraphics[width=0.48\textwidth]{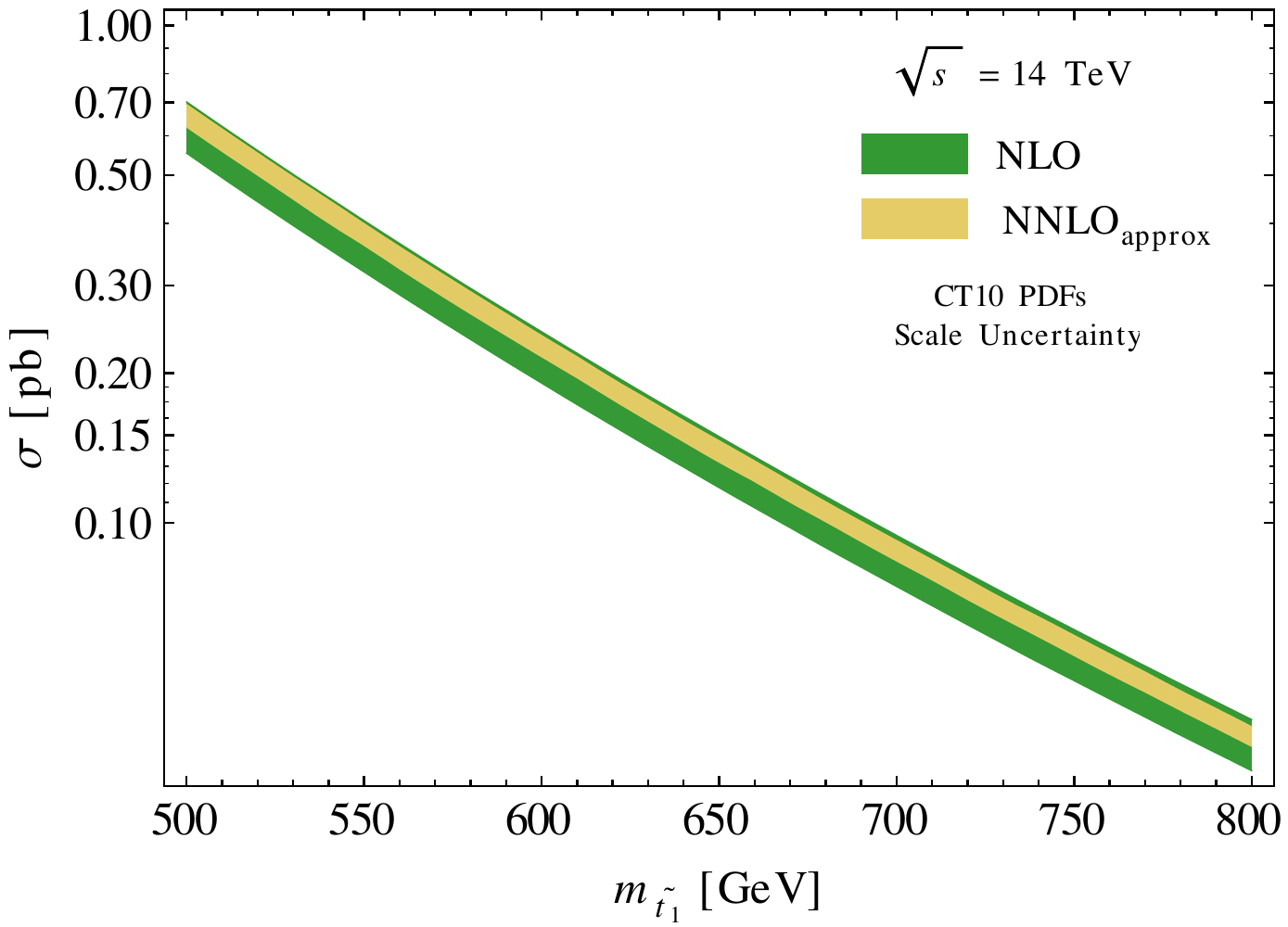} & \includegraphics[width=0.48\textwidth]{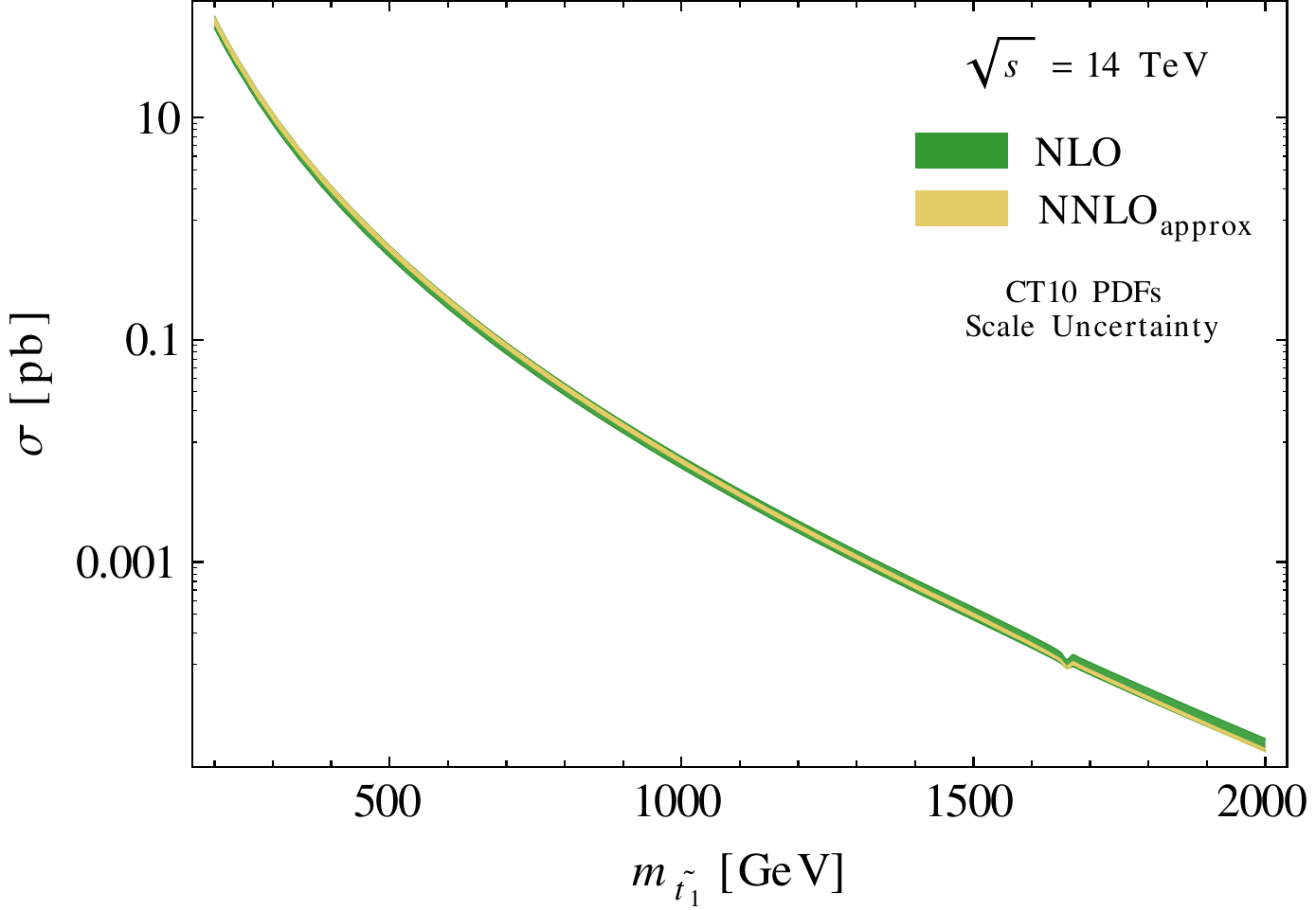} \\ 
\end{tabular} 
\end{center}
\caption{Mass scans with CT10 PDFs for the LHC with $\sqrt{S}=7$, $8$, and $14$~TeV. The bands represent the perturbative scale uncertainties at NLO and NNLO. \label{fig:CT10massscan}}
\end{figure}

\begin{figure}[t]
\begin{center}
\begin{tabular}{cc}
 \includegraphics[width=0.48\textwidth]{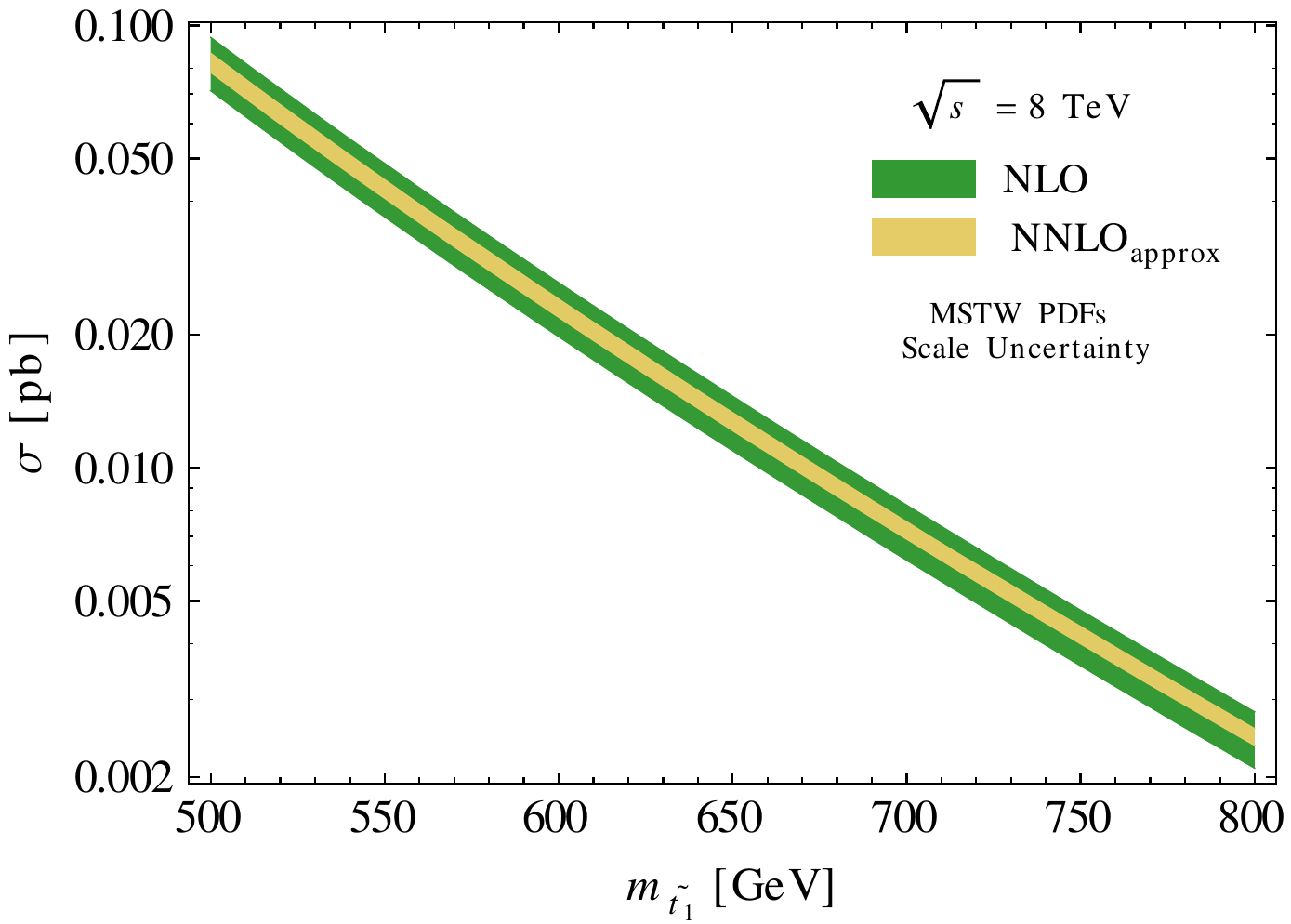} & \includegraphics[width=0.48\textwidth]{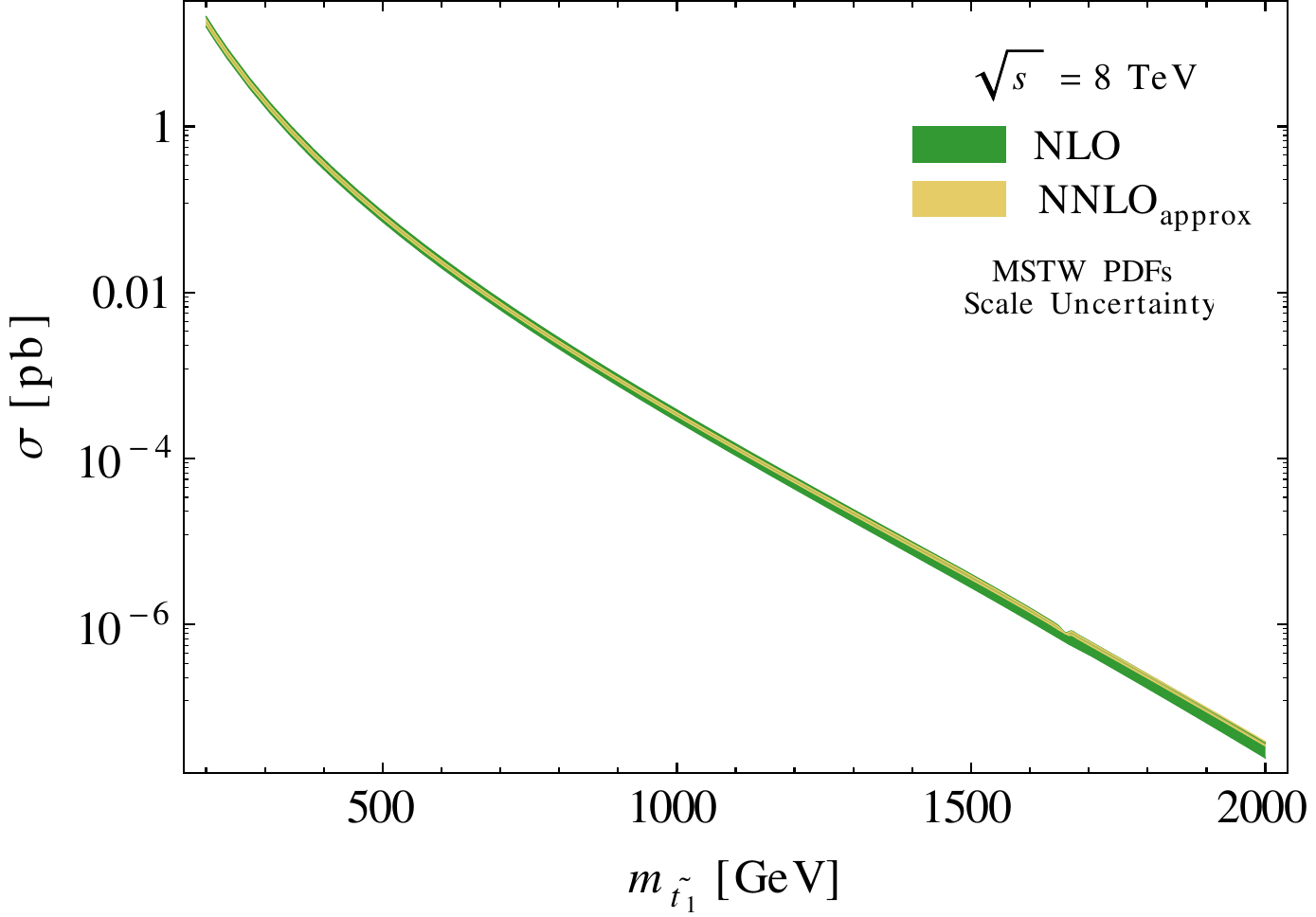} \\ 
\end{tabular} 
\end{center}
\caption{Mass scans with MSTW2008 PDFs for the LHC at $\sqrt{S}=8$~TeV. The bands represent the perturbative scale uncertainties at NLO and NNLO. \label{fig:MSTWmassscan}}
\end{figure}

Finally,  Figure~\ref{fig:CT10massscan}  shows the cross section as a function of the top-squark mass up to $m_{\tilde{t}_1} = 2$~TeV for the LHC at $7$, $8$ and $14$~TeV center of mass energy. In all cases, the bands represent the residual perturbative uncertainties, obtained as explained above. In Figure~\ref{fig:CT10massscan} we employ CT10 PDFs. One sees that, for large values of the stop mass,  the approximate NNLO band is below the NLO band  at $7$ and $8$~TeV center of mass energy, while, for the LHC at $14$~TeV, the approximate NNLO band overlaps with the  lower part of the NLO band.
For comparison, in Figure~\ref{fig:MSTWmassscan} we repeat the same analysis by employing MSTW2008 PDFs for the LHC at $8$~TeV. In this case, for large stop masses the approximate NNLO scale uncertainty bands tend to be slightly above the NLO bands.

\begin{table}[t]
\begin{center}
\begin{tabular}{|c||c|c|}
\hline \rule[-2ex]{0pt}{5.5ex} CTEQ6.6 PDFs& $m_{\tilde{t}_1} = 100$~GeV & $m_{\tilde{t}_1} = 400$~GeV 
\\ 
\hline
\hline \rule[-2ex]{0pt}{5.5ex} $(\sigma \pm \Delta \sigma_\mu)_{\mbox{{\footnotesize NLO+NLL}}}$  [pb] from \cite{Beenakker:2010nq} & $390^{+53}_{-41}$  & $0.209^{+0.018}_{-0.019} $    \\ 
\hline \rule[-2ex]{0pt}{5.5ex} $(\sigma \pm \Delta \sigma_\mu)_{\mbox{{\footnotesize approx. NNLO}}}$ [pb], this work  &  $393^{+39}_{-22}$ &  $0.215^{+0.010}_{-0.012}$  \\ 
\hline 
\end{tabular} 
\end{center}
\caption{Comparison between the NLO+NLL results of \cite{Beenakker:2010nq} and the approximate NNLO results of the present work. The table refers to the LHC  with  $\sqrt{S} = 7$~TeV; the PDFs employed are from CTEQ6.6 \cite{Nadolsky:2008zw}. We report only the perturbative uncertainties. \label{fig:compNLL7}}
\end{table}

\begin{table}[t]
\begin{center}
\begin{tabular}{|c||c|c|}
\hline \rule[-2ex]{0pt}{5.5ex} CTEQ6.6 PDFs& $m_{\tilde{t}_1} = 100$~GeV & $m_{\tilde{t}_1} = 400$~GeV 
\\ 
\hline
\hline \rule[-2ex]{0pt}{5.5ex}  $(\sigma \pm \Delta \sigma_\mu)_{\mbox{{\footnotesize NLO+NLL}}}$  [pb] from  \cite{Beenakker:2010nq}& $1.65^{+0.22}_{-0.16} \times 10^3$  & $2.19^{+0.20}_{-0.19}$    \\ 
\hline \rule[-2ex]{0pt}{5.5ex}    $(\sigma \pm \Delta \sigma_\mu)_{\mbox{{\footnotesize approx. NNLO}}}$  [pb], this work 
& $1.64^{+0.17}_{-0.08} \times 10^3$ &  $2.22^{+0.13}_{-0.10}$   \\ 
\hline 
\end{tabular} 
\end{center}
\caption{Comparison between the NLO+NLL results of \cite{Beenakker:2010nq} and the approximate NNLO results of the present work. The table refers to the LHC  with  $\sqrt{S} = 14$~TeV; the PDFs employed are from CTEQ6.6. We report only the perturbative uncertainties. \label{fig:compNLL14}}
\end{table}

We conclude this Section by comparing our results with the ones presented in \cite{Beenakker:2010nq} and \cite{Falgari:2012hx}, which have NLO+NLL accuracy. In Tables~\ref{fig:compNLL7} and \ref{fig:compNLL14} we show the results obtained for the input parameters employed in \cite{Beenakker:2010nq}, which coincide with the SPS1a' benchmark scenario of \cite{AguilarSaavedra:2005pw},  as well as for two different values of the stop mass. Table~\ref{fig:compNLL7} refers to the LHC with 
$7$~TeV center of mass energy while Table~\ref{fig:compNLL14} refers to the LHC with a center of mass energy of $14$~TeV.
We checked that, as expected, the NLO results in \cite{Beenakker:2010nq} coincide with the ones obtained with our modified version of {\tt Prospino}. The central values of our approximate NNLO predictions for the total cross section are in very good agreement with the ones obtained by means of NLO+NLL formulas in \cite{Beenakker:2010nq}. However, the perturbative uncertainties at approximate NNLO accuracy appear to be smaller that the NLO+NLL scale uncertainties.

\begin{table}[t]
\begin{center}
\begin{tabular}{|c||c|c|c|c|c|}
\hline $m_{\st}$ [GeV]  & NLO  [pb] & NLL \cite{Falgari:2012hx} [pb] &  $\text{NNLO}_{\text{approx}}$ [pb]
& $\hat{K}_{\text{NLL}}$ \cite{Falgari:2012hx} & 
$\hat{K}_{\text{NNLO}_{\text{approx}}}$ 
 \\ 
 \hline
\hline $100$ & $415^{+64}_{-59}$ & $477^{+87}_{-66}$ & $425^{+45}_{-25}$ &  $1.14$ & $1.02$  \\ 
\hline $200$ & $12.7^{+1.8}_{-1.8}$  & $14.7^{+2.5}_{-1.9}$  & $13.3^{+1.0}_{-0.8}$ & $1.15$  & $1.04$ \\ 
\hline $300$ & $1.27^{+0.17}_{-0.19}$ & $1.49^{+0.24}_{-0.19}$ & $1.35^{+0.08}_{-0.09}$ & $1.17$ & $1.06$\\ 
\hline $400$ & $0.211^{+0.028}_{-0.031}$ & $0.250^{+0.038}_{-0.030}$ & $0.226^{+0.011}_{-0.014}$ & $1.18$ & $1.07$ \\ 
\hline $800$ & $1.09^{+0.16}_{-0.18}\times 10^{-3}$ & $1.34^{+0.20}_{-0.16}\times 10^{-3}$ & $1.22^{+0.04}_{-0.08}\times 10^{-3}$ & $1.23$ & $1.12$ \\ 
\hline $1000$ & $1.24^{+0.19}_{-0.21}\times 10^{-4}$ & $1.57^{+0.23}_{-0.18}\times 10^{-4}$ & $1.42^{+0.04}_{-0.1}\times 10^{-4}$ & $1.27$ & $1.15$ \\ 
\hline 
\end{tabular} 
\end{center}
\caption{Comparison between the NLO+NLL results of \cite{Falgari:2012hx} and the approximate NNLO results of the present work.  The numbers refer to the benchmark point {\tt 40.2.4} \cite{AbdusSalam:2011fc}. In particular, we set $m_t = 172.5$~GeV, $m_{\tilde{g}}=1386$~GeV, $m_{\tilde{q}}=m_{\tilde{t}_2}=1358$~GeV and $\cos \alpha=0.39$  as in \cite{Falgari:2012hx}. The factorization scale is set equal to $m_{\st}$. For the approximate NNLO results we used MSTW2008 NLO PDFs. It must be observed that the numbers in \cite{Falgari:2012hx} were obtained with a private version of the MSTW NLO PDFs.\label{fig:compNLLref}}
\end{table}


In \cite{Falgari:2012hx}, the authors perform a simultaneous resummation of the production threshold logarithms and Coulomb singularities at  NLL accuracy, including bound-state effects. Table 7 in \cite{Falgari:2012hx} reports result for the production of top squark pairs at the CMSSM benchmark point {\tt 40.2.4} \cite{AbdusSalam:2011fc}. Results for higher values of the stop quark mass can be found in the files provided with the arXiv submission of \cite{Falgari:2012hx}.
 Coulomb resummation 
and bound state effects increase the cross section, but the largest effect in the NLL results of  \cite{Falgari:2012hx} is due to soft resummation. The authors of  \cite{Falgari:2012hx} employ a private version 
of the MSTW2008 NLO PDFs in their calculations.
In Table~\ref{fig:compNLLref} we compare our findings with the ones of~\cite{Falgari:2012hx}. One can immediately see
that our cross section prediction (inclusive of perturbative uncertainty)  has a sizable  overlap with the predictions
quoted in \cite{Falgari:2012hx} (inclusive of perturbative uncertainty). In all cases, the central value of our predictions falls within the perturbative uncertainty of the NLL calculations.
The central values of the approximate NNLO calculations are lower than the ones of  \cite{Falgari:2012hx}, as it can be seen from the fact that the quantities  $\hat{K}_i \equiv \sigma_i/\sigma_{\text{NLO}}$ are lower than the corresponding quantities in \cite{Falgari:2012hx}. 
In analyzing this fact it is necessary  to take into account that our result does not include the additional enhancements due to Coulomb corrections, bound state effects and resummation of scale-dependent logarithms in the hard function\footnote{In our set-up, the latter would correspond to keep the contributions of the scale-dependent terms in ${\bm H}_{ij}^{(2)}$, see Eq.~(\ref{eq:NNLOapp})}. In this sense, it would be more natural to compare our results with the predictions labeled $\text{NLL}_{s}$ and $\text{NLL}_{s,\rm fixed} $ in \cite{Falgari:2012hx}; these two approaches give $\hat{K}$ factors which are roughly $0.03$-$0.08$ smaller than the ones corresponding to the NLL predictions (see Fig.~15 in \cite{Falgari:2012hx}), and fully compatible with our results.
Additionally, experience with top quark production indicates that  NNLL resummation gives cross sections which are larger than the ones obtained from the corresponding approximate NNLO formulas \cite{Ahrens:2010zv, Ahrens:2011mw}.

However, at this point in our discussion, the stage is set to comment on the reciprocal advantages and disadvantages of the various methods employed to implement soft gluon resummation  and/or  approximate NNLO formulas. This point can be better illustrated in the case of top quark pair production, where full NNLO results recently became available
\cite{Baernreuther:2012ws, Czakon:2013goa}. The various calculational schemes adopted  by different groups
can be classified with respect to two aspects:
\emph{i)} The kind of kinematics employed, and \emph{ii)} the space in which the resummation is carried out. Many predictions for the total top quark pair cross section are obtained by employing the traditional production threshold kinematics identified by the $\beta \to 0$ limit \cite{Aliev:2010zk, Beneke:2011mq},
while PIM kinematics and 1PI kinematics, which additionally allow one to calculate differential distributions of phenomenological interest in top quark physics,  are employed in \cite{Ahrens:2010zv} and \cite{Kidonakis:2010dk, Ahrens:2011mw}, respectively. One can then decide to work in Mellin moment space \cite{Aliev:2010zk, Kidonakis:2010dk}, or in directly in momentum space \cite{Beneke:2011mq,Ahrens:2010zv, Ahrens:2011mw}.
All of these schemes allow one to obtain NNLL or approximate NNLO accuracy and they differ in the kind of formally subleading terms which are neglected. Without guidance from complete NNLO calculations there are no definitive arguments to prefer a priori one scheme to another. What one can do is to see how a given approximation does at NLO, where exact results are known, along the lines we followed at the beginning of this section. For the top quark production, all of the predictions obtained in the various schemes are compatible with each other when perturbative uncertainties are taken into account \cite{Kidonakis:2013zsa}. It is only a posteriori, when the results of a full NNLO calculation are available, that one can see which of the various approximate NNLO/NNLL results is closest to the exact NNLO result. In the case of the top quark total cross section, our method, based on PIM and 1PI kinematics and momentum space, provides results which are slightly lower than, but compatible with, the results of the other groups. It is important to stress that there is not a clear pattern indicating, even a posteriori, a definite methodological bias: the approximate NNLO prediction which best approximates the full NNLO total cross section is based upon 1PI kinematics and Mellin space
\cite{Kidonakis:2010dk}, while Mellin space and momentum space calculations in the production threshold limit provide very similar results \cite{Aliev:2010zk, Beneke:2011mq}. A detailed analysis of the various approaches was carried out in Section 5.2 of \cite{Ahrens:2011mw}; in particular, it was shown that the approach based on SCET and employed  here significantly improves the agreement between the results obtained in 1PI and PIM kinematics. Furthermore, it is possible to argue that the good numerical agreement between the exact NLO result at the LHC and the corresponding approximation based on production threshold expansion is somewhat fortuitous:
 In fact, in spite of the fact that  the $\beta \to 0$ expansion fails to reproduce the correct shape of the 
(unphysical) distribution $d \sigma/ d \beta$ in the gluon fusion channel, the integral of the approximate distribution is very close to the full NLO result (see for example the discussion in \cite{Ahrens:2010mj}). 
On the other hand it is fair to say that the difference between the full NNLO calculation of the top production cross section and 
the results of \cite{Ahrens:2010zv,Ahrens:2011mw, Ahrens:2011px}  is an a posteriori indication that the neglected formally subleading terms in those works turned out to be numerically larger than the neglected subleading terms in the other approaches.

In conclusion, we believe that, in absence of a full NNLO calculation,
approaches leading to different kinds of approximate NNLO /NNLL predictions are well worth exploring.
Furthermore,  numerical differences among predictions based on different approaches should be conservatively taken as a measure of the uncertainty associated to the neglected subleading corrections. Precisely for this reason,  following the procedure of \cite{Ahrens:2011px}, we decided to base our predictions for the stop production cross section on an average of the PIM and 1PI kinematics calculations and we estimated the perturbative error not only by means of the scale variation, but also by considering  the maximal difference between the predictions obtained in the two kinematic schemes, as explained in this section.


\section{Conclusions}
\label{sec:Conc}

Supersymmetry is certainly one of the best motivated scenarios 
for physics beyond the standard model. If supersymmetry is broken just above 
the electroweak scale, the supersymmetric partners of the known 
particles are expected to have masses of the order of 1 TeV, 
and they could soon be observed at the LHC. For these reasons it 
is important to have precise theoretical predictions for the production 
cross sections of supersymmetric particles at hadron colliders.
Top squarks are expected to be among the lightest colored supersymmetric 
particles, and consequently they might be the supersymmetric partners which are most easily 
accessible at the LHC. In this 
paper we have employed effective field theory techniques in order to improve existing 
NLO calculations for stop-pair production by evaluating higher-order perturbative 
corrections arising from soft-gluon emissions. 

In particular, we have adopted a framework which allows one to resum large logarithmic corrections at NNLL accuracy in the stop pair production process. The resummation can be carried out  in two different
kinematic schemes, referred to as PIM and 1PI. In principle, this fact enables one to obtain predictions for both the pair invariant mass distribution and the top-squark $p_T$ and rapidity distributions, as well as for the total cross section, which is at the moment the observable of primary interest in phenomenological studies. Furthermore,  by re-expanding the NNLL resummation  formulas, 
it is possible to obtain predictions for the cross section 
which have approximate NNLO accuracy in fixed-order perturbation theory. 
The evaluation of these approximate NNLO formulas represents the main goal of this work, where we obtained analytic expressions for all of the coefficients 
multiplying the plus distributions in the variables $(1 - z)$ and $s_4$  in the NNLO hard-scattering kernels for the double differential distributions
in the two kinematic schemes considered. This was made possible by a complete computation of the hard-function matrices in both production channels at NLO accuracy, the calculation of which represents the main technical result of the present work.

The impact of the approximate NNLO corrections on the stop pair production cross section has been examined.
In order to obtain the best possible predictions and a better control on the neglected subleading terms, 
we have averaged 
the approximate NNLO results obtained in PIM and 1PI kinematics and matched them with exact fixed-order NLO calculations. As in the case of slepton-pair production studied in \cite{Broggio:2011bd}, 
we found that the total cross section 
depends strongly on the mass of the produced particles, while
the dependence on the remaining SUSY  parameters, such as the masses of other supersymmetric particles, is rather weak. 
We have found that including the approximate NNLO predictions for the pair production 
cross section reduces the perturbative uncertainty 
by more than a factor of two with respect  to the NLO results. We stress that we explicitly accounted for the kinematic 
scheme uncertainty arising from the use of approximate NNLO formulas and we combined it with the scale uncertainty, as explained in 
Section~\ref{sec:CS}.
It was found that the approximate  NNLO corrections have only a moderate impact on the central 
value of the NLO total cross section. 
We compared our result with analogous calculations at NLO+NLL accuracy \cite{Beenakker:2010nq,Falgari:2012hx} which were carried out by means of methods  which are different from the ones employed in this work and within different kinematic schemes. We found very good agreement with the results of \cite{Beenakker:2010nq}  and a substantial agreement within the quoted perturbative uncertainties with \cite{Falgari:2012hx}, 
although the cross section values we found tend to be slightly smaller than the ones quoted in \cite{Falgari:2012hx}.
We conclude by observing that, after the NNLO corrections are included, the main theoretical 
uncertainty on the total cross section comes from the PDFs. This is particular evident when large values of the stop 
mass are considered. Therefore,  the results presented in this paper allow one to improve 
the precision of the stop-pair total cross section  predictions by reducing their scale dependence
to the extent that it becomes  considerably smaller than PDF uncertainty.

\section*{Acknowledgments}

We would like to thank A.~von Manteuffel and S.~Berge for useful discussions and T.~Plehn for assistance with {\tt{Prospino}}.
The work of A.F.\ was supported in part by the PSC-CUNY Award No.\ 65214-00-43 and by the National Science Foundation Grant No.\ PHY-1068317. The research of M.N.\ is
supported by the ERC Advanced Grant EFT4LHC of the European Research Council, the
Cluster of Excellence {\emph{Precision Physics, Fundamental Interactions and Structure of Matter}}
(PRISMA -- EXC 1098) and grant NE 398/3-1 of the German Research Foundation (DFG),
grants 05H09UME and 05H12UME of the German Federal Ministry for Education and Research (BMBF), and the Rhineland-Palatinate Research Center {\emph{Elementary Forces and Mathematical Foundations}}. L.V.\ acknowledges the Alexander von Humboldt Foundation for support.

\appendix

\section{NLO soft functions \label{SFA}}

In this appendix we collect the results of the calculation of the NLO soft function in PIM kinematics \cite{Ahrens:2010zv} and in 1PI kinematics \cite{Ahrens:2011mw}.

It was proven in \cite{Becher:2007ty} that the Laplace-transformed soft function defined in Eq.~(\ref{eq:SoftLaplace}) is related to the position-space Wilson loop \cite{Korchemsky:1993uz}.
In particular, the soft function in position space is defined as (see Eq.~(39) in \cite{Ahrens:2010zv})
\begin{align}
\bm{W}(x,\mu) = \frac{1}{d_R} \langle 0 | \bar{\bm{T}} [\bm{O}^{s \dagger}(x)] 
\bm{T}[ \bm{O}^{s}(0)] | 0 \rangle \, ,
\end{align}
where $\bm{T}$ ($\bar{\bm{T}}$) indicates time ordering (anti-time ordering), while $\bm{O}_s$ is the operator
\begin{align}
\bm{O}_s(x) =  [\bm{S}_n \bm{S}_{\bar{n}} \bm{S}_{v_3} \bm{S}_{v_4}] (x) \, .
\label{eq:Os}
\end{align}
The four soft Wilson lines   $\bm{S}$ are oriented along the directions of the incoming partons ($n$ and $\bar{n}$), and along the four-velocities of the outgoing top squarks ($v_3$ and $v_4$). The precise definition of the soft Wilson lines can be found in Eqs.~(22) in \cite{Ahrens:2010zv}. The Laplace transformed soft functions can be obtained from the soft functions in position space through the relation
\begin{align}
\tilde{\bm{s}}\left(L,\mu\right) 
& = \bm{W}\left(x_0 = \frac{-2i}{e^{\gamma_E} \mu e^{L/2}}, \mu \right) . \label{eq:SoftLap}
\end{align}
Here and below, we have omitted the dependence of the soft functions on the PIM or 1PI kinematic variables and on the heavy particle masses, as well as the subscripts $q \bar{q}$ or $gg$ indicating the channel.
The expansion of $\tilde{\bm{s}}$ in powers of the strong coupling constant is 
\be
\tilde{\bm{s}} = \tilde{\bm{s}}^{(0)} + \frac{\alpha_s}{4 \pi} \tilde{\bm{s}}^{(1)} + \left(\frac{\alpha_s}{4 \pi}\right)^2 \tilde{\bm{s}}^{(2)} + {\mathcal O}(\alpha_s^3)
\, . 
\ee
At leading order the soft functions are the same both in PIM and 1PI kinematics:
\be
\tilde{\bm{s}}_{q \bar{q}}^{(0)} = \left(
\begin{matrix}
N & 0 \\
0 & \frac{C_F}{2}
\end{matrix} \right) , \qquad
\tilde{\bm{s}}_{gg}^{(0)} = \left(
\begin{matrix}
N & 0 &0 \\
0 & \frac{N}{2} & 0 \\
0 & 0 & \frac{N^2-4}{2 N}
\end{matrix} \right) .
\ee

The bare soft function at NLO in position space can be written as 
\be
\bm{W}^{(1, i)}_{{\rm bare}} (\epsilon, x_0, \mu) = \sum_{ij} \bm{w}_{ij} {\mathcal I}^{k}_{ij}(\epsilon, x_0, \mu) \, , \quad (k = {\rm PIM}, {\rm 1PI}) \, .
\ee
The matrices $\bm{w}_{ij}$ are related to the products of color generators and are the same for both kinematics. In the quark annihilation channel they are
\begin{align}
  \label{eq:qTT}
  \bm{w}_{12}^{q\bar{q}} = \bm{w}_{34}^{q\bar{q}} &= -\frac{C_F}{4N}
  \begin{pmatrix}
    4N^2 & 0
    \\
    0 & -1
  \end{pmatrix}
  , \nonumber
  \\
  \bm{w}_{33}^{q\bar{q}} = \bm{w}_{44}^{q\bar{q}} &= \frac{C_F}{2}
  \begin{pmatrix}
    2N & 0
    \\
    0 & C_F
  \end{pmatrix}
  , \nonumber
  \\
 \bm{w}_{13}^{q\bar{q}} = \bm{w}_{24}^{q\bar{q}}
 &=- \frac{C_F}{2}
  \begin{pmatrix}
    0 & 1
    \\
    1 & 2C_F - \frac{N}{2}
  \end{pmatrix}
  , \nonumber
  \\
  \bm{w}_{14}^{q\bar{q}} = \bm{w}_{23}^{q\bar{q}} &= -\frac{C_F}{2N}
  \begin{pmatrix}
    0 & -N
    \\
    -N & 1
  \end{pmatrix}
  ,
\end{align}
while for the gluon fusion channel one finds
\begin{align}
  \label{eq:gTT}
  \bm{w}_{12}^{gg}&=- \frac{1}{4}
  \begin{pmatrix}
    4N^2 & 0 & 0
    \\
    0 & N^2 & 0
    \\
    0 & 0 & N^2-4
  \end{pmatrix}
  , \nonumber
  \\
  \bm{w}_{34}^{gg}&=-
  \begin{pmatrix}
    C_FN & 0 & 0
    \\
    0 & -\frac{1}{4} & 0
    \\
    0 & 0 & -\frac{N^2-4}{4N^2}
  \end{pmatrix}
  , \nonumber
  \\
  \bm{w}_{33}^{gg}=\bm{w}_{44}^{gg}&= \frac{C_F}{2N}
  \begin{pmatrix}
    2N^2 & 0 & 0
    \\
    0 & N^2 & 0
    \\
    0 & 0 & N^2-4
  \end{pmatrix}
  , \nonumber
  \\
 \bm{w}_{13}^{gg}= \bm{w}_{24}^{gg}&= -\frac{1}{8}
  \begin{pmatrix}
    0 & 4N & 0
    \\
    4N & N^2 & N^2-4
    \\
    0 & N^2-4 & N^2-4
  \end{pmatrix}
  , \nonumber
  \\
  \bm{w}_{14}^{gg}= \bm{w}_{23}^{gg}&= -\frac{1}{8}
  \begin{pmatrix}
    0 & -4N & 0
    \\
    -4N & N^2 & -(N^2-4)
    \\
    0 & -(N^2-4) & N^2-4
  \end{pmatrix}
  .
\end{align}

The functions  ${\mathcal I}^{i}_{ij}$ are integrals over the soft-gluon phase-space. In PIM kinematics one finds ${\mathcal I}^{{\rm PIM}}_{11} = {\mathcal I}^{{\rm PIM}}_{22}  = 0$ and 
\begin{align}
  \mathcal{I}^{{\rm PIM}}_{12} &= -\left( \frac{2}{\epsilon^2} + \frac{2}{\epsilon} L_0 + L_0^2 +
    \frac{\pi^2}{6}\right) , \nonumber
  \\
  \mathcal{I}^{{\rm PIM}}_{33} = \mathcal{I}^{{\rm PIM}}_{44} &= \frac{2}{\epsilon} + 2L_0 - \frac{2}{\beta_t} \ln
  x_s \, , \nonumber
  \\[1mm]
  \mathcal{I}^{{\rm PIM}} _{34} &= -\frac{1+x_s^2}{1-x_s^2} \left[ \left( \frac{2}{\epsilon} + 2L_0
    \right) \ln x_s - \ln^2x_s + 4\ln x_s \ln(1-x_s) + 4\Li_2(x_s) - \frac{2\pi^2}{3}
  \right] , \nonumber
  \\[1mm]
  \mathcal{I}^{{\rm PIM}} _{13} = \mathcal{I}^{{\rm PIM}}_{24} &= -\left[\frac{1}{2} \left( L_0 -
      \ln\frac{(1+y_t)^2x_s}{(1+x_s)^2} \right)^2 + \frac{\pi^2}{12} +
    2\Li_2\left(\frac{1-x_s y_t}{1+x_s}\right) +
    2\Li_2\left(\frac{x_s-y_t}{1+x_s}\right)\right] , \nonumber
  \\[2mm]
  \mathcal{I}^{{\rm PIM}}_{14} = \mathcal{I}^{{\rm PIM}}_{23} &= \mathcal{I}_{13}(y_t \to z_u) \, ,
\end{align}
where $\beta_t = \sqrt{1 - 4 m_{\st}^2/M^2}$ coincides with $\beta$ in the $z \to 1$ limit.
Furthermore $x_s=(1-\beta_t)/(1+\beta_t)$, $y_t=-t_1/m_{\st}^2-1$, $z_u=-u_1/m_{\st}^2-1$, and
\begin{align}
  L_0 = \ln \bigg( -\frac{\mu^2x_0^2e^{2\gamma_E}}{4} \bigg) \,.
\end{align}
 In 1PI kinematics one finds again that ${\mathcal I}^{{\rm 1PI}}_{11} = {\mathcal I}^{{\rm 1PI}}_{22}  = 0$, while 
\begin{align}
  \mathcal{I}^{{\rm 1PI}}_{12} &= - \left[ \frac{2}{\epsilon^2} + \frac{2}{\epsilon} \left( L_0 -
      \ln\frac{s' m_{\st}^2}{t_1u_1} \right) + \left( L_0 -
      \ln\frac{s' m_{\st}^2}{t_1u_1} \right)^2 + \frac{\pi^2}{6} + 2\Li_2
    \left( 1 - \frac{s' m_{\st}^2}{t_1u_1} \right) \!\right] , \nonumber
  \\
  \mathcal{I}^{{\rm 1PI}}_{33} &= \frac{2}{\epsilon} + 2L_0 - \frac{2(1+\beta_t^2)}{\beta_t} \ln x_s
  \, , \nonumber
  \\
  \mathcal{I}^{{\rm 1PI}}_{44} &= \frac{2}{\epsilon} + 2L_0 + 4 \, , \nonumber
  \\
  \mathcal{I}^{{\rm 1PI}}_{14} = \mathcal{I}^{{\rm 1PI}}_{24} &= -\frac{1}{\epsilon^2} - \frac{1}{\epsilon} L_0
  - \frac{1}{2} L_0^2 - \frac{\pi^2}{12} \, ,
  \\
  \mathcal{I}^{{\rm 1PI}}_{13} &= - \Biggl[ \frac{1}{\epsilon^2} + \frac{1}{\epsilon} \left(L_0 - 2
    \ln{\frac{t_1}{u_1}} \right) + \frac{1}{2} \left( L_0 -
    2\ln\frac{t_1}{u_1} \right)^2 + \frac{\pi^2}{12} \nonumber
  \\
  &\hspace{1.1cm} + 2\Li_2 \left( 1 - \frac{t_1}{u_1x_s} \right) + 2\Li_2
  \left( 1 - \frac{t_1x_s}{u_1} \right) \Biggr] \, , \nonumber
  \\
  \mathcal{I}^{{\rm 1PI}}_{23} &= \mathcal{I}'_{13} \, (t_1 \leftrightarrow u_1) \, ,
  \nonumber
  \\
  \mathcal{I}^{{\rm 1PI}}_{34} &= \frac{1+\beta_t^2}{2\beta_t} \left[- \frac{2}{\epsilon} \ln x_s
    -2L_0\ln x_s + 2\ln^2x_s - 4\ln x_s \ln(1-x_s^2) - 2\Li_2(x_s^2) + \frac{\pi^2}{3}
  \right] .\nonumber
\end{align}
In the case of 1PI kinematics, the definition of $\beta_t$  should be changed to $\beta_t = \sqrt{1- 4 m_{\st}^2/s'}$.

\bibliography{paperbib060312}
\bibliographystyle{JHEP-2}

\end{document}